

\input epsf.tex

\font\rmu=cmr10 scaled\magstephalf
\font\bfu=cmbx10 scaled\magstephalf

\font\it=cmti10 scaled \magstephalf
\font\bf=cmbx10 scaled\magstephalf
\rmu

\font\rmus=cmr8
\font\rmuss=cmr6

\font\mait=cmmi10 scaled\magstephalf
\font\maits=cmmi7 scaled\magstephalf
\font\maitss=cmmi7

\font\msyb=cmsy10 scaled\magstephalf
\font\msybs=cmsy8 scaled\magstephalf
\font\msybss=cmsy7

\font\bfus=cmbx7 scaled\magstephalf
\font\bfuss=cmbx7

\font\cmeq=cmex10 scaled\magstephalf

\textfont0=\rmu
\scriptfont0=\rmus
\scriptscriptfont0=\rmuss

\textfont1=\mait
\scriptfont1=\maits
\scriptscriptfont1=\maitss

\textfont2=\msyb
\scriptfont2=\msybs
\scriptscriptfont2=\msybss

\textfont3=\cmeq
\scriptfont3=\cmeq
\scriptscriptfont3=\cmeq

\newfam\bmufam  \textfont\bmufam=\bfu
      \scriptfont\bmufam=\bfus \scriptscriptfont\bmufam=\bfuss

\hsize=15.5cm
\vsize=22cm
\baselineskip=16pt   
\parskip=16pt plus  2pt minus 2pt

\def\ni{\noindent}

\def\lf{\leaders\hbox to 1em{\hss.\hss}\hfill}

\def\a{\alpha}
\def\b{\beta}
\def\d{\delta}
\def\e{\epsilon}
\def\g{\gamma}
\def\l{\lambda}

\def\del #1{\frac{\partial^{#1}}{\partial\l^{#1}}}

\def\R{{\rm I\!R}}
\def\one{{\mathchoice {\rm 1\mskip-4mu l} {\rm 1\mskip-4mu l}
{\rm 1\mskip-4.5mu l} {\rm 1\mskip-5mu l}}}
\def\C{{\mathchoice
{\setbox0=\hbox{$\displaystyle\rm C$}\hbox{\hbox to0pt
{\kern0.4\wd0\vrule height0.9\ht0\hss}\box0}}
{\setbox0=\hbox{$\textstyle\rm C$}\hbox{\hbox to0pt
{\kern0.4\wd0\vrule height0.9\ht0\hss}\box0}}
{\setbox0=\hbox{$\scriptstyle\rm C$}\hbox{\hbox to0pt
{\kern0.4\wd0\vrule height0.9\ht0\hss}\box0}}
{\setbox0=\hbox{$\scriptscriptstyle\rm C$}\hbox{\hbox to0pt
{\kern0.4\wd0\vrule height0.9\ht0\hss}\box0}}}}

\font\fivesans=cmss10 at 4.61pt
\font\sevensans=cmss10 at 6.81pt
\font\tensans=cmss10
\newfam\sansfam
\textfont\sansfam=\tensans\scriptfont\sansfam=\sevensans\scriptscriptfont
\sansfam=\fivesans
\def\sans{\fam\sansfam\tensans}
\def\Z{{\mathchoice
{\hbox{$\sans\textstyle Z\kern-0.4em Z$}}
{\hbox{$\sans\textstyle Z\kern-0.4em Z$}}
{\hbox{$\sans\scriptstyle Z\kern-0.3em Z$}}
{\hbox{$\sans\scriptscriptstyle Z\kern-0.2em Z$}}}}
\def\del{\partial}

\newcount\foot
\foot=1
\def\note#1{\footnote{${}^{\number\foot}$}{\ftn #1}\advance\foot by 1}

\def\frac#1#2{{#1\over #2}}

\font\ch=cmbx12 scaled\magstephalf
\font\ftn=cmr8 scaled\magstephalf

\font\it=cmti10 scaled\magstephalf
\font\bf=cmbx10 scaled\magstephalf

\font\titch=cmbx12 scaled\magstep2
\font\titname=cmr10 scaled\magstep2
\font\titit=cmti10 scaled\magstep1
\font\titbf=cmbx10 scaled\magstep2

\nopagenumbers


\line{\hfil CGPG-93/9-1}
\line{\hfil Sept 8, 1993}
\epsffile{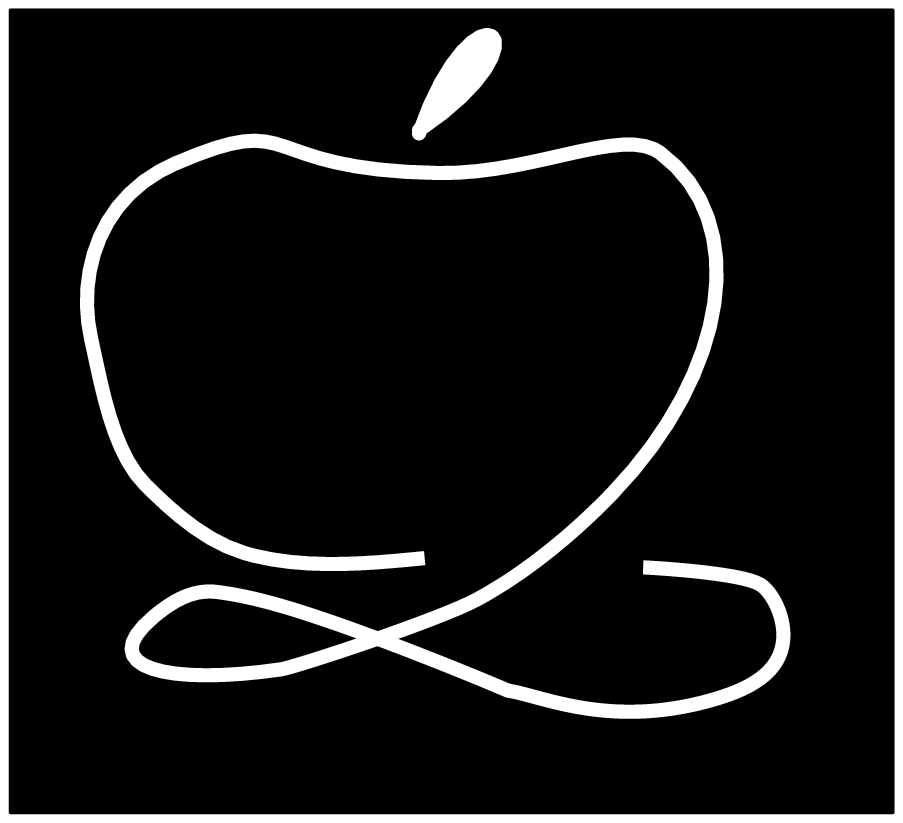}
\vskip.5cm
\centerline{\titch CHROMODYNAMICS AND GRAVITY}
\vskip.5cm
\centerline{\titch AS THEORIES ON LOOP SPACE}
\vskip1.2cm
\centerline{\titname R. Loll }
\vskip.5cm
\centerline{\titit Center for Gravitational Physics and Geometry}
\vskip.2cm
\centerline{\titit Pennsylvania State University}
\vskip.2cm
\centerline{\titit University Park, PA 16802-6300}
\vskip.2cm
\centerline{\titit U.S.A.}

\vskip1.5cm
\centerline{\titbf Abstract.\hfill}
\ni
We review some attempts of reformulating both gauge theory and general
relativity in terms of holonomy-dependent loop variables. The emphasis
lies on exhibiting the underlying mathematical structures, which often
are not given due attention in physical applications. An extensive list
of references is included.

\vfill\eject



\epsffile{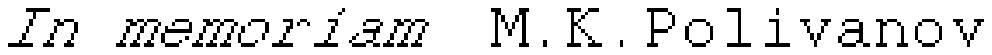}
\vfill\eject
\footline={\hss\tenrm\folio\hss}
\pageno=1

\line{\ch Contents\hfil}
\vskip.6cm

\line{ 1. Introduction\lf p.1}
\line{ 2. Paths and loops\lf p.2}
\line{ 3. Holonomy\lf p.3}
\line{ 4. Wilson loops\lf p.6}
\line{ 5. The space ${\cal A}/\cal G$\lf p.7}
\line{ 6. The space of all loops and its topology\lf p.9}
\line{ 7. Structures on loop space\lf p.11}
\line{ 8. Identities satisfied by (traced) holonomies\lf p.16}
\line{ 9. Equivalence between gauge potentials and holonomies\lf p.20}
\line{10. Physical interpretation of holonomy\lf p.25}
\line{11. Classical loop equations\lf p.27}
\line{12. Differential operators\lf p.31}
\line{13. Lattice gauge theory\lf p.35}
\line{14. Loop algebras\lf p.38}
\line{15. Canonical quantization\lf p.41}
\line{\hskip.65cm References\lf p.45}

\vskip2cm

\line{\ch 1. Introduction \hfil}

This paper is an introduction to the use of loop variables in gauge theory
and gravity, and to some of its underlying mathematical structures. It is a
considerably enlarged version of a previous review paper on loop approaches
to gauge field theory [Lol7]. Its main mathematical ingredient is the
holonomy, i.e. the integral along an open or closed path of a gauge potential
in space(-time).

Although this review is restricted to the application of loop methods to
Yang-Mills theory and general relativity (in the Ashtekar formulation) in
four space-time dimensions, many of the mathematical issues discussed in the
initial sections (2-9) are relevant to any theory whose basic configuration
variable is a connection one-form, as are, for example, some of the so-called
topological
field theories. The remaining sections (10-15) deal more specifically with
the physical applications. I describe and summarize the status of
path- and loop-dependent formulations of Yang-Mills theory and gravity,
focussing on a few selected topics.

One of my aims is to illustrate that there is not just one, but many
different loop spaces, and that the physical properties of a theory
should be taken into consideration when making a specific choice. Another
aim is to try and explain why past attempts of using loop variables have
often run into problems, in the hope that this may aid further research in
this direction. A large number of references is dicussed in the text whose
selection however reflects my own bias and does not claim to be complete.
For additional references and complementary expositions, the reader may
consult the review sections of [Gu, Mig2, Bar].

\vskip1.5cm

\line{\ch 2. Paths and loops \hfil}

Given a differentiable, simply connected manifold $\Sigma$ of dimension
$d$, a {\it path} in $\Sigma$ (Fig.1a) is a continuous map $w$ from a closed
interval of the real line $\R$ into $\Sigma$,

$$
\eqalign{w:&[s_1,s_2]\rightarrow\Sigma\cr
               & s\;\mapsto\; w^{\mu}(s).\cr}\eqno(2.1)
$$

\ni As such it has the properties of a map between two differentiable
manifolds, for example, ($C^r$) differentiability, piecewise
differentiability or non-differentiability, and its tangent vector
$\frac{dw}{ds}$ may vanish for some or all parameter values $s$.

A {\it loop} in $\Sigma$ (Fig.1b) is a closed path, by which we shall mean a
continuous map $\gamma$ of the unit interval into $\Sigma$,

$$
\eqalign{\g:&[0,1]\rightarrow\Sigma\cr
              &s\;\mapsto\;\g^{\mu}(s),\cr
               \g&(0)=\g(1).\cr}\eqno(2.2)
$$

\ni We will be using such closed paths in the construction of
gauge-invariant quantities in pure gauge theory. Open paths play an
important role in gauge theory with fermions, where natural
gauge-invariant objects are open flux lines with quarks or Higgs fields
``glued to
the endpoints" [KogSus, GaGiTr, ForGam, GamSet].

\epsffile[0 10 460 220]{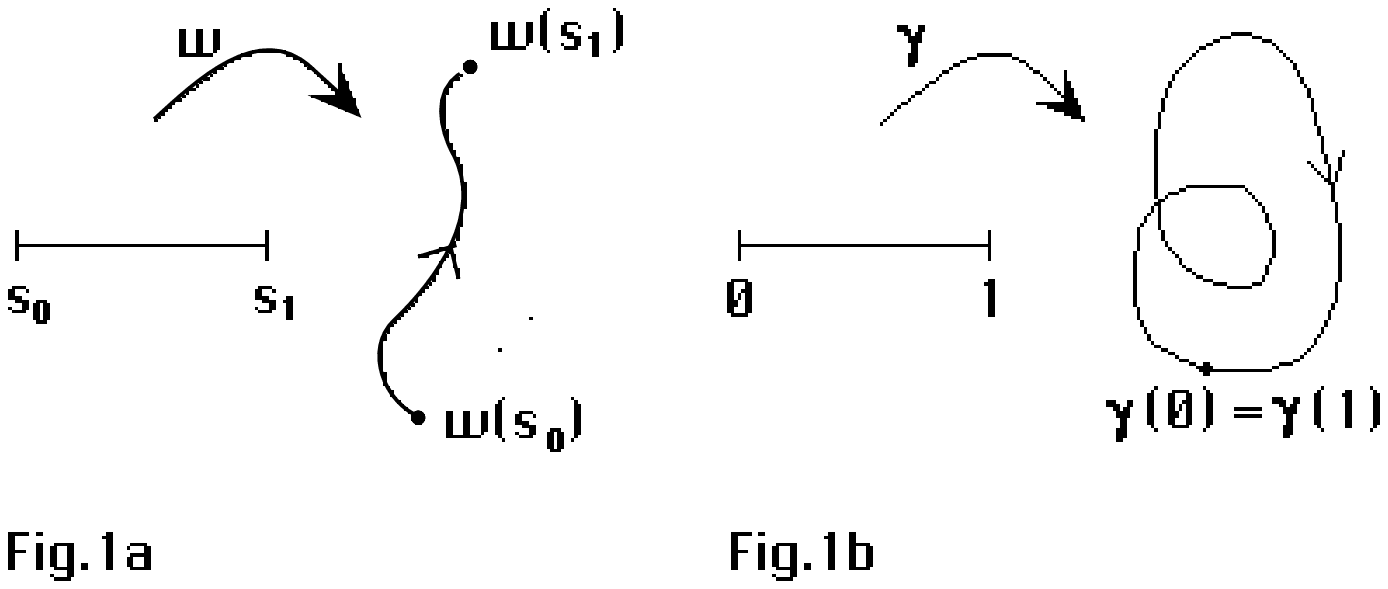}

The manifold $\Sigma$ may be the real vector space $\R^d$, possibly with
a Euclidean or Minkow\-skian metric, but
may also be non-linear and topologically non-trivial. In this case the
possibility arises of having non-contractible loops, i.e. maps $\g$
that cannot be continuously shrunk to a {\it point loop}

$$
p_x(s)=x\in \Sigma,\quad\forall s\in [0,1].\eqno(2.3)
$$

\ni Note that even if a path is closed it has a distinguished image point,
namely, its initial and end point, $\g(0)=\g(1)$, and that each point in
the image of $\g$ is labelled by one or more (if there are selfintersections)
parameter values $s$.

\vskip1.5cm

\line{\ch 3. Holonomy\hfil}

Suppose in an open neighbourhood $V$ of $\Sigma$ we are given a configuration
$A\in \cal A$, the set of all {\it gauge
potentials} $\Sigma\rightarrow\Lambda^1${\bf g}, i.e. a smooth
{\bf g}-valued connection one-form, with
{\bf g} denoting the Lie algebra of a finite-dimensional Lie
group $G$. We have

$$
A(x)=A_{\mu}(x)\, dx^{\mu}= A_{\mu}^a(x) X_a\,dx^\mu, \eqno(3.1)
$$

\ni where $X_a$ are the algebra generators in the fundamental representation
of {\bf g} ($a=1\dots$ dim $G$) and $x^{\mu}$, $\mu=1\dots d$, a set of
local coordinates on $V$.

The {\it holonomy} $U_w$ of a path $w^\mu (s)$ with initial point $s_0$ and
endpoint $s_1$ (whose image is completely contained in $V$) is the solution
of the system of differential equations

$$
\frac{dU_w(s,s_0)}{ds}= A_{\mu}(x)\frac{dw^{\mu}}{ds} U_w(s,s_0),\quad
 s_0\leq s\leq s_1,\eqno(3.2)
$$

\ni with $x=w(s)$, subject to the initial condition

$$
U_w(s_0,s_0)=e,\eqno(3.3)
$$

\ni where $e$ denotes the unit element in $G$. Note that this
definition only makes sense for at least piecewise differentiable paths
$w$. The solution of (3.2) is given by the {\it path-ordered exponential
of $A$ along $w$},

$$
\eqalign{&U_w(s_1,s_0)=\,{\rm P}\; \exp\; g{\int_{s_0}^{s_1} A_{\mu}(w(t))
\dot w^{\mu}(t) dt}:=\cr
&:=\one + \sum_{n=1}^\infty g^n \int_{s_0}^{s_1} dt_1 \int_{t_1}^{s_1}
dt_2\dots \int^{s_1}_{t_{n-1}}dt_n\, A_{\mu_n}(w(t_n)) \dots A_{\mu_1}
(w(t_1))\dot w^{\mu_n}(t_n)\dots \dot w^{\mu_1}(t_1).\cr}\eqno(3.4)
$$

\ni The coupling constant $g$ is necessary to render the argument of the
exponential dimensionless. Note that from (3.4) follows the composition
law $U_w(s_2,s_0)=U_w(s_2,s_1)U_w(s_1,s_0)$, for $s_0\leq s_1
\leq s_2$.
An alternative definition for $U_w$ that does not need differentiability of
the path and employs an approximation of $w$ by $n$ straight line segments
$(x_i-x_{i-1})$ is as the limit

$$
U_w(s_1,s_0)=\lim_{n\rightarrow\infty}
(1+A(x_n)\,(x_n-x_{n-1}))
(1+A(x_{n-1})\,(x_{n-1}-x_{n-2}))
\dots (1+A(x_1)\,(x_1-x_0)),
\eqno(3.5)
$$

\ni with $\sup ||x_i -x_{i-1}||\rightarrow 0$ as $n$ increases [CorHas], and
where $x_0=w(s_0),\, x_1,\dots$, $x_n=w(s_1)$ is a set of $n+1$ points ordered
along the path $w$ (Fig.2).

\epsffile[0 50 490 280]{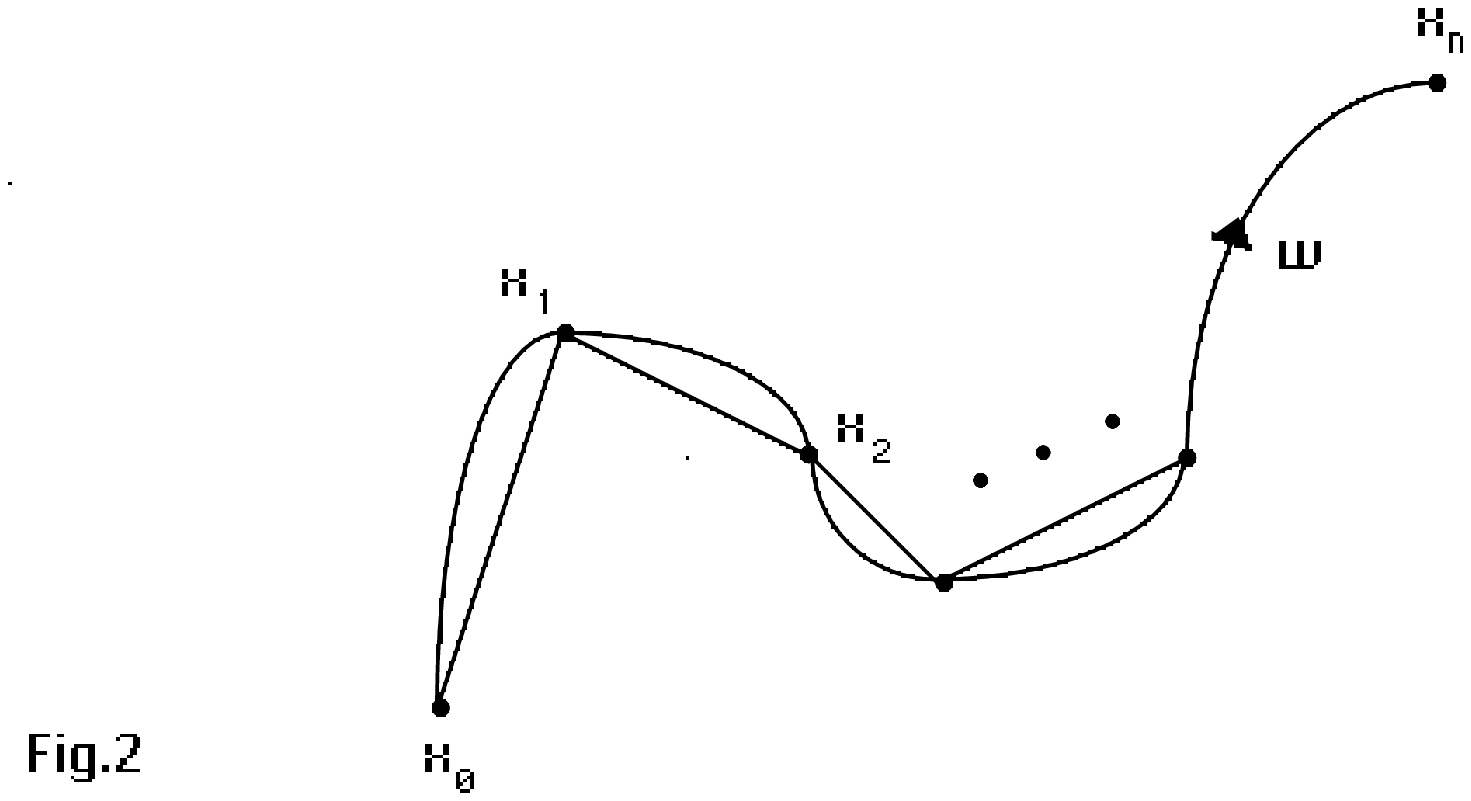}

The holonomy $U_w$ takes its values in $G$ and transforms under
gauge transformations (smooth functions $g:V\rightarrow G$)
according to

$$
U_w(s_1,s_0){\buildrel g\over \longmapsto}g^{-1}(w(s_1))U_w(s_1,s_0)g(w(s_0)).
\eqno(3.6)
$$

\ni Note this would not be true if we had allowed for discontinuities of
the path $w$. The corresponding change of the gauge potential is
straightforwardly computed from equation (3.2),

$$
A_\mu (x){\buildrel g\over\longmapsto} g^{-1}(x) A_\mu (x)g(x)-
\frac{dg^{-1}(x)}{dx^\mu}g(x)= g^{-1}(x) A_\mu (x)g(x)+g^{-1}(x)
\frac{dg(x)}{dx^\mu}.\eqno(3.7)
$$

\ni Another property of $U_w(s_1,s_0)$ following from (3.2) is its invariance
under smooth orientation-preserving reparametrizations $f$ of $w$, i.e.

$$
w(s)=w'(f(s))\Longrightarrow U_w(s_1,s_0)=U_{w'}(f(s_1),f(s_0)),\eqno(3.8)
$$

\ni where $\frac{df}{ds}>0$, $\forall s$. The term ``non-integrable"
(i.e. path-dependent) ``phase factor" for $U_w$ was introduced by Yang
[Yan], in generalization of its abelian version for $U(1)$-electromagnetism.
It is a well-known result in mathematics that the connection
$A$ can (up to gauge transformations) be reconstructed from the
knowledge of the holonomies of the {\it closed} curves based at a
point $x_0\in\Sigma$ (see [KobNom, Lic] for details on the concept
of holonomy group and related issues). The reader may consult the paper by
Barrett [Bar] for an account of the mathematical development of these
so-called reconstruction theorems, and for more references. The details of
such theorems depend on the mathematical setting, for instance whether the
underlying fibre bundle is differentiable or just topological. Also
generalizations of the concept of connection are possible.

The above description is valid only in a local neighbourhood $V$ of $\Sigma$.
The appropriate global description is afforded by the mathematical theory
of the principal fibre bundles $P: G\rightarrow P\rightarrow\Sigma$
over the base manifold $\Sigma$ with typical fibre $G$, with
$A^P$ a connection one-form on $P$. Typically $P$ has no global cross
sections (this depends both on the group $G$ and the manifold
$\Sigma$), and then $A^P$ can be identified with the one-form (3.1) only
in a coordinate patch $V\subset\Sigma$, using a local cross section. The
global implications of this fact are well known (see, for example, the
discussion in [ChoDeW]) and will not be addressed in this paper.
Suffice it to remark that the holonomy variables are global in the sense
that because of their gauge-invariance they do not ``see" non-trivial
transition functions between coordinate patches on the manifold $\Sigma$.
For more comments on global issues, see for example [Bar, Fis, HoScTs, Gro].
However, there are important examples where global cross sections do exist
(for instance, $G=SU(2)$ and any three-dimensional $\Sigma$) and hence
all results described here are valid globally.

Note that for $F=0$ (also called the case of a ``flat connection"), where
$F$ is the field strength tensor,

$$
F_{\mu\nu} (x)=\del_\mu A_\nu (x)-\del_\nu A_\mu (x)+ig [A_\mu (x),A_\nu (x)],
\eqno(3.9)
$$

\ni $U_w$ depends only on the homotopy of the path $w$.

\vskip1.5cm

\line{\ch 4. Wilson loops\hfil}

In this section we will be working exclusively with loops, i.e. closed
paths $\g$ in $\Sigma$, and their associated holonomies,

$$
U_\g= {\rm P}\exp \oint_{\g}A_\mu(x) dx^\mu.\eqno(4.1)
$$

For a loop $\g$ based at a point
$x_0\in\Sigma$ there is a natural way of constructing a gauge-invariant
quantity, namely,

$$
T_A(\g):= \frac{1}{N}{\rm Tr}\, U_\g= \frac{1}{N}{\rm Tr}\,{\rm P}\exp
\oint_{\g}A_\mu(x) dx^\mu, \eqno(4.2)
$$

\ni the {\it traced holonomy} or so-called {\it Wilson loop} (introduced by
Wilson [Wil] as an indicator of confining behaviour in lattice gauge
theory). The trace is taken in a linear representation $R$ of the group
$G$, and the normalization factor in front depends on the dimension $N$
of $R$. Under a gauge transformation, $U_{\g,x_0}$ transforms homogeneously
according to (3.5), but the matrices $g(x_0)$ and $g^{-1}(x_0)$ cancel each
other upon taking the trace. Because of the cyclicity of the trace $T_A(\g)$ is
independent of the choice of base point $x_0=\g(0)=\g(1)$ of $\g$. Note that
$T_A(\g)$ is invariant under orientation-preserving reparametrizations of $\g$
and is therefore a function on {\it unparametrized loops}, i.e.
reparametrization equivalence classes $\bar\g$, where any two members of
$\bar\g$ are related by a smooth orientation-preserving bijection $b:S^1\mapsto
S^1$, $\frac{db}{ds}>0$. Hence an unparametrized loop can be thought of as a
continuous ordered set of points in $\Sigma$, with a (positive or negative)
orientation assigned to each point in a continuous manner (as long as $\g$ does
not have any selfintersections or is a point loop (2.3)).

An unparametrized loop possesses certain (differential) topological
properties (i.e. invariants under smooth diffeomorphisms of $\Sigma$),
such as its number of selfintersection points, number of points of
non-differentiability (``kinks"), its winding number and its knot class
[Kau1,2]. This feature is important for the application of loop methods to
gravitational theories, as will be explained in more detail below. In concrete
calculations involving an unparametrized loop $\bar\g$ one usually works with a
chosen member $\g$ from the equivalence class $\bar\g$ and then ensures the end
result is independent of this choice. There have also been attempts to set up a
loop calculus that is intrinsically reparametrization-invariant [GamTri3,4,
Mig2, Gam, Tav].

\vskip1.5cm

\line{\ch 5. The space ${\cal A}/\cal G$\hfil}

There is a more global, geometric way of looking at the setting described in
the last section. We have assumed that the configuration space of the theories
under consideration is a space $\cal A$ of {\bf g}-valued connection one-forms,
on
which the infinite-dimensional group $\cal G$ of local gauge transformations
acts according to (3.7). The physical configuration space therefore consists of
the orbits in $\cal A$ under this gauge group action, where different elements
$A(x)$ of the same orbit are gauge-related and hence physically
indistinguishable. This quotient space is usually denoted by ${\cal A}/\cal
G$, but more precisely it may have two different meanings.
The first one is the ``covariant quotient" where $\cal A$
consists of space-time connections, and the group $\cal G$ of local gauge
transformations on space-time, the other one is the quotient that comes up in
the Hamiltonian formulation (in the $A_0=0$-gauge), where both the connections
and the gauge transformations are now purely spatial.

In any case, the quotient ${\cal A}/\cal G$ of Yang-Mills theory is a
non-linear
space with a non-trivial topology, and many of its properties as an
infinite-dimensional differentiable manifold are well known. (Strictly
speaking, one has to remove some singular points from $\cal A$ (those where
the connection is reducible), to make the quotient space into a manifold.)
One of the main problems of gauge theory has been to find a suitable set of
variables to describe this quotient space and the physical dynamics in an
efficient way, and which at the same time could serve as a starting point for a
non-perturbative quantization scheme. In order to get rid of at least part of
the non-trivial transformation behaviour of the gauge potentials $A$, the
non-local holonomy-dependent variables discussed here are a natural choice.

However, it has to be remembered that if one starts from some linear space of
loop or path functions, there have to be non-linear constraints if a subset of
these functions is to serve as ``coordinates" (in the
strict mathematical sense) on the space ${\cal A}/\cal G$. In fact, the
essence of these non-local reformulations is to find necessary and sufficient
conditions on the elements of such a function space which ensure they are
in one-to-one correspondence with the equivalence classes $[A]\in{\cal A}/\cal
G$. Usually this goes under the name of ``reconstruction theorems", which were
already mentioned earlier, and will be discussed in more detail in Sec.9.

In the application to the Ashtekar formulation of canonical gravity in 2+1 and
3+1 dimensions, it does not suffice to form the quotient of the relevant
spaces of connections $\cal A$ with respect to the groups of local
$SO(2,1)$- and $SU(2)_\C$-gauge rotations respectively. One still has to
factor out by the diffeomorphisms, which also correspond to gauge symmetries,
i.e. the gauge equivalence classes $[A]$ do {\it not} correspond to
observables in general relativity. Nevertheless loop-dependent variables
$f(\g)$ have
turned out to be useful in this context, because they carry a non-trivial
representation of the diffeomorphism group $Diff(\Sigma)$ of the spatial
manifold $\Sigma$, namely,

$$
(\phi f)(\g):=f(\phi\circ\g),\qquad \forall\phi\in {\it Diff}(\Sigma),
\eqno(5.1)
$$

\ni with $\circ$ denoting the composition of maps. Invariance under spatial
diffeomorphisms is therefore expressed by the condition

$$
\phi f=f,\qquad \forall\phi\in {\it Diff}(\Sigma).\eqno(5.2)
$$

\ni Similar conditions are often adopted to express the diffeomorphism
invariance of loop-depen\-dent quantum operators. - There is yet another new
feature that occurs in the application to gravity: since the gauge group
$SU(2)_\C=SL(2,\C)$ (and also $SU(1,1)$) is non-compact, the quotient
${\cal A}/\cal G$ is in fact non-Hausdorff, i.e. its singularities are more
ill-behaved than in the compact case. The implications of this result are
discussed in [AshLew1].

In case one considers the set of loops based at a point $x_0$ in a
manifold $\Sigma$, the gauge dependence of their associated holonomies (4.1)
is reduced from the infinite-dimensional group $\cal G$ to the
finite-dimensional group of gauge transformations $G$ at the point $x_0$. This
is sometimes rephrased by saying that the set of holonomies based at a point
$x_0$ is invariant under the restricted gauge group ${\cal G}_{x_0}$,
consisting of those elements of $\cal G$ which act trivially at the base point.
By contrast, the Wilson loops (4.2) are invariant with respect to the entire
group $\cal G$, and therefore are functions on the quotient space
${\cal A}/\cal G$.

Alternative descriptions for the space ${\cal A}/\cal G$ or an appropriate
generalization thereof are also of interest in the quantum theory, where this
space appears as the integration domain in the ``sum over all configurations"
of
the path integral approach and the scalar product of the Hamiltonian approach
(see [Lol6] for an introduction). An attempt to use an abelian $C^*$-algebra
generated by the Wilson loop variables
for this purpose is described in [AshIsh1], and developed further in
[AshLew2, Bae].

\vskip1.5cm

\line{\ch 6. The space of all loops and its topology\hfil}

Since it will be of relevance to the field theoretic application later,
let us try to give a mathematically meaningful definition of the
``space of all loops". In mathematics, the {\it loop space} $\Omega_{x_0} X$
associated with some topological space $X$ with distinguished base point
$x_0$ is usually taken to be the function space

$$
\Omega_{x_0}X=(X,x_0,x_0)^{({\rm I},0,1)}\eqno(6.1)
$$

\ni of continuous functions $\g:$ I $\mapsto X$ from the unit interval
I $=[0,1]$ to $X$ such that $\g(0)=\g(1)=x_0$ [Ada]. In most of the following,
we will omit the explicit reference to the base point $x_0$.

In order to have a notion of convergence for sequences of points
in $\Omega X$ (i.e. sequences $\g_i$ of functions, $i=0,1,2,\dots$)
and a notion of continuity for functions $f:\Omega X\rightarrow Y$
into some space $Y$, one has to give $\Omega X$ a topological
structure. A standard choice [Ada, Mic] is the compact-open topology
in which open sets are given by subsets of $\Omega X$ of the form

$$
\Phi(J,O):=\{\g |\g(J)\subset O,\hbox{ $J$ a closed interval in $[0,1]$,
$O$ an open set in $X$}\},\eqno(6.2)
$$

\ni together with their unions and finite intersections [Lip]. It is
straightforward to show that with this topology $\Omega X$ is
Hausdorff (if we assume $X$ to be Hausdorff), i.e. distinct points in
$\Omega X$ possess disjoint neighbourhoods. However, as was pointed out by
Barrett [Bar], holonomy mappings are not in general continuous in the
compact-open topology, which therefore seems not suitable for physical
applications where holonomies play a central role. He also emphasizes the role
that ought to be played by physical measurements in the choice of a topology.
Unfortunately, neither for non-abelian gauge theory nor for gravity this is
particularly straightforward. It does however seem reasonable to make a
choice in which gauge-invariant observables
rather than gauge-covariant quantities are
continuous. To this end
one may employ a well-known construction, namely to induce a topology on loop
space by demanding that a given set of functions on it be continuous. Barrett
suggests to use the group-valued holonomies as a preferred
set of functions. In this context a function $f:\Omega X\rightarrow\R$ is
continuous if for every one-parameter family $\g_t$ of loops the composition
map $f(\g_t)$ is a continuous function of $t$. The same induced topology is
used by Lewandowski [Lew]. Similarly, one may demand that the
complex-valued Wilson loop variables (4.2) be continuous.

There is another possibility for defining a topology on loop space, if one
uses a Riemannian metric $g$ on the underlying manifold $\Sigma$. It is induced
by the distance $d$ on loop space, where

$$
d(\g_1,\g_2)=\inf_S Area(S),\eqno(6.3)
$$

\ni and $S$ runs through all two-surfaces that are bounded by the loop
$\g_1\circ\g_2^{-1}$. This gives rise to a well-defined metric on loop space
(see [DurLei] for more details). It is a natural definition
since it takes into account the thin equivalence of loops (cf. Sec.8) in that
two thinly equivalent loops have vanishing distance $d$.

Note that $\Omega X$
is not in general a vector space since we cannot add elements of
$\Omega X$. However, if $X$ is a linear space and if we choose
$x_0=0$, we can obtain a linear structure on $\Omega X$ by defining
addition and scalar multiplication pointwise:

$$
\eqalign{(\g_1+\g_2)(s)&:=\g_1(s)+\g_2(s)\cr
(a\g)(s)&:= a\cdot\g(s),\quad a\in\R\,(\C),\cr}\eqno(6.4)
$$

\ni (the dot denoting scalar multiplication in $X$), which because of the
continuity of these operations makes $\Omega X$ into a topological
vector space. If $X$ is a differential manifold $\Sigma$ and we
consider $\Omega\Sigma$ as consisting only of differentiable maps $\g$,
$\Omega\Sigma$ can be made into an infinite-dimensional differential manifold
(see [Mic] for a discussion of natural topologies on spaces of continuous
and differentiable mappings). The finest topology one can impose on a loop
space is the discrete topology, in which any element of $\Omega\Sigma$ is
an open set. It has been used by Ashtekar and Isham in their treatment
of representations of loop algebras [AshIsh1, Ish].

If for some reason one does not want to distinguish a base point in $X$, a
natural loop space to work with is

$$
{\cal L}X:=\bigcup_{x\in X} \Omega_x X,\eqno(6.5)
$$

\ni with $\Omega_x X$ as defined in (6.1). - In physical applications one is
usually interested in subspaces or quotient spaces of $\Omega X$ or
${\cal L}X$. Examples of the former are restrictions to the sets of loops
without selfintersections, loops without kinks, or contractible loops.
Typical quotient spaces are those of loops modulo orientation-preserving
reparametrizations, or loops modulo constant translations as employed by
Mensky [Men1,2,3] for the special case of $X=\R^n$. In these cases one
has to check
which properties of the original loop space (and the physical dynamics
defined in terms of $\Omega X$ or ${\cal L}X$) are compatible with the
restriction or the projection to the quotient space, respectively.

\vskip1.5cm

\line{\ch 7. Structures on loop space\hfil}

Let me emphasize that assigning well-defined mathematical properties to
loop space is not a superfluous luxury but necessary if one
wants to set up a meaningful differential calculus on $\Omega X$ (cf. Sec.12
below).
{}From a physical point of view it is of eminent importance to have additional
(topological, algebraic, ...) structure defined on $\Omega X$, which is
preserved (or approximately preserved) in the quantum theory.

A loop space $\Omega X$ is ``better" than an arbitrary infinite-dimensional
topological space because its elements can be composed,
with the product map given by

$$
(\g_1\circ\g_2)(s)=\cases  {\g_1(2s), &$0\leq s\leq\frac12$\cr
                          \g_2(2s-1) &$\frac12\leq s\leq 1$.\cr}\eqno(7.1)
$$

\ni This product is neither commutative nor associative, and we have neither
a unit nor inverse loops in $\Omega X$. (If we work with the loop space
${\cal L}X$, then composition is only defined if the endpoint of $\g_1$
coincides with the initial point of $\g_2$.) Taking the inverse $\g^{-1}$ of
a loop $\g$ to be

$$
\g^{-1}(s):=\g(1-s),\quad\forall s,\eqno(7.2)
$$

\ni defines an involution on $\Omega X$ since $(\g^{-1})^{-1}=
\g$, and $(\g_1\circ\g_2)^{-1}=\g_2{}^{-1}\circ\g_1{}^{-1}$.
Note that modifying loop space to

$$
\bar\Omega_{x_0}X:=\bigcup_{\rm I}(X,x_0,x_0)^{([0,b],0,b)},\eqno(7.3)
$$

\ni with the union extending over all closed intervals I$=[0,b]$,
$b\geq 0$, and the product map defined by

$$
(\g_1\circ\g_2)(s)=\cases {\g_1(s) &$0\leq s\leq b_1$\cr
                           \g_2(s) &$b_1\leq s\leq b_1+b_2$,\cr}\eqno(7.4)
$$

\ni which is associative (though not commutative), and with the unit given by
the trivial loop $\g_0:[0,0]\mapsto x_0$, $\bar\Omega X$ becomes a
topological semigroup (see [Sta] for more comments on associativity, in the
context of string theory).

One way $\Omega X$ may inherit structure is from $X$, for example,
from a Riemannian metric on $X$ (see below).
Another case we will be concerned
with is when a quotient space $\Omega X /\sim$ acquires an
algebraic structure via certain functions defined on $\Omega\Sigma$, as
for example the Wilson loop function (see Sec.8).

There is a limited amount of rigorous mathematical results on spaces of
unparametrized loops, some of which will be summarized in the following.
Both Brylinski [Bry] and Sch\"aper [Sch1,2] describe the space of
unparametrized
loops as the base space of a principal {\it Diff}$^+ (S^1)$\-bundle. Since the
quotient space ${\cal L}\Sigma/Diff^+(S^1)$ is singular, they choose as
starting
point the subset $X=E(S^1,\Sigma)\subset {\cal L}\Sigma$ of smooth embeddings
of
the circle into the Riemannian manifold $\Sigma$.
(For $\dim \Sigma \geq 3$, $X$ is a dense open subspace of $C^\infty
(S^1,\Sigma)$, consisting of those elements $\gamma\in C^\infty
(S^1,\Sigma)$ for which $s\not= t$ (mod $1$) implies
$\gamma(s)\not=\gamma(t)$. This excludes all loops with
self-intersections and retracings, and the trivial point loops.)
Equipping it with the usual $C^\infty$-topology, one thus obtains a Fr\'echet
principal fibre bundle $X(Y,Diff^+(S^1))$ ($Y$ denoting the quotient manifold
$Y=E(S^1,\Sigma)/Diff^+(S^1)$), where $X$ and $Y$ are modelled on the spaces
$C^\infty(S^1,\R^d)$ and $C^\infty(S^1,\R^{d-1})$, respectively.

Moreover, this construction extends to the larger manifold $\hat X$ of singular
loops (of which $X$ is a dense open subspace), defined as the immersions
$S^1\to \Sigma$ with finitely many self-intersections and tangency points, each
of finite order [Bry], and there is an analogous principal fibre bundle $\hat
X(\hat Y, Diff^+(S^1))$.
For $dim\,\Sigma=3$, and given a Riemannian metric $g$ on $\Sigma$,  there is a
natural almost-complex structure $J$ on $\hat Y$. If $v$ is a tangent vector to
$\hat Y$, and $\vec e$ the unit tangent vector to the loop $\g\in \hat X$, then
for every point $p$ of the corresponding $\tilde\g\in \hat Y$, define

$$
J\cdot v(p):=\vec e \times v(p).\eqno(7.5)
$$

\ni There is also a natural weak symplectic structure on $\hat Y$, induced from
the $Diff^+(S^1)$-invariant two-form

$$
\b_\g(u,v)=\int_0^1\nu_{\g(s)}(\dot\g(s),u(s),v(s))ds\eqno(7.6)
$$

\ni on ${\cal L}\Sigma$, where $\nu$ denotes a nowhere vanishing volume form on
$\Sigma$. The corresponding form $\b$ on $\hat Y$ is non-degenerate in the weak
sense that it induces an injection on $T_\g\hat Y\to T^*_\g\hat Y$ with dense
image. Hence the Poisson brackets are not defined for all smooth functions, but
only on the subset of so-called super-smooth functions. Finally, there is a
natural Riemannian metric structure on $\hat Y$, given by

$$
G(v,w)=\int_0^1 g(v(s),w(s))||\dot\g(s)||ds.\eqno(7.7)
$$

\ni where $||\dot\gamma(s)||$ denotes the norm of the
tangent vector to the loop $\gamma$ at $s$. Its relation with the symplectic
and complex structures is $G(v,w)=\b(v,J\cdot w)$.

Sch\"aper calculates the Levi-Civita connection, the Riemannian curvature
tensor and the sectional curvature on the Riemannian manifold $(X,G)$.
Furthermore, the metric $G$ gives rise
to a natural connection one-form on the principal bundle
$X\rightarrow Y$. Some explicit solutions
for horizontal geodesics (corresponding to trajectories of free motion of
unparametrized loops) on $X$ are given for the
special case of $\Sigma=\R^3$ [Sch1].
Note that this construction is not $Diff(\Sigma)$-invariant,
because it makes explicit use of the metric $g$ on $\Sigma$.

Slightly different settings are explored in two papers by Fulp.
In the first one [Ful1], a smooth action of the Fr\'echet Lie group
$Diff(I)$ on the space ${\cal P}^*\Sigma$ of non-degenerate paths  (elements
$\gamma$ of ${\cal P}\Sigma=C^\infty (I,\Sigma)$ with nowhere vanishing tangent
vector $\dot\gamma(s)$) on a manifold $\Sigma$ is defined such that
${\cal P}^*\Sigma/Diff(I)$ inherits a Fr\'echet manifold structure.
Smooth vector fields and one-forms are defined as smooth sections of the
bundles $T({\cal P}^*\Sigma)\to {\cal P}^*\Sigma$ and $T^*({\cal P}^*\Sigma)\to
{\cal P}^*\Sigma$, respectively. It is then proven that, given a smooth
$(k+1)$-form $\tilde\a$ on $\Sigma$, there is a smooth differential form $\a$
on
${\cal P}^*\Sigma$ defined by

$$
\a(\d_1,\dots,\d_k):=\int_0^1\tilde\a_{\g(t)}(\dot\g(t),\d_1(t),\dots,\d_k(t)),
\eqno(7.8)
$$

\ni where $\d_i\in T_\g({\cal P}^*\Sigma)$. These so-called path forms have
appeared (in a more formal treatment) in the physics literature, for example,
in [CoqPil]. It is not clear how they
relate to general differential forms on ${\cal P}^*\Sigma$. Explicit formulae
for the exterior path space derivative of certain path forms are given, for
example, that of the path-ordered exponential ${\cal I}: {\cal P}^*P\to G$,
${\cal I}(\g)={\rm P}\,\exp\int_\g\omega$.

Finally, Fulp derives an identity, to be thought of as ``the
analogue of the Yang-Mills operator on loop space", essentially equating the
integral of the classical Yang-Mills equations along a closed path in $P$ with
a certain differential operator on loop space, linear in $\a$, where $\a$ is
the pullback ${\cal I}^*(\theta)$ to ${\cal P}^*P$ of the Maurer-Cartan form
$\theta$ on $G$. It would be interesting to see if this expression can serve as
a starting point of a classical reformulation of Yang-Mills theory on loop
space.

In a second paper [Ful2], Fulp investigates natural structures on
the principal bundle
${\cal P}P({\cal P}\Sigma, C^\infty (I,G))$, the path space of a
principal bundle $P(\Sigma,G)$, and various subspaces of ${\cal P}P$ in
a Fr\'echet manifold setting. He mainly deals with the
submanifold ${\cal P}_{u_0}P\subset {\cal P}P$ of paths with a fixed
base point $u_0\in P$. It is shown that the corresponding
principal bundle ${\cal P}_{u_0}P({\cal P}_{\pi(u_0)}\Sigma, C^\infty_0
(I,G))$ is trivial. The Fr\'echet Lie group $C_0^\infty(I,G)$ is
the subgroup of all elements $a\in C^\infty(I,G)$ such that
$a(0)=e$, with $e$ denoting the unit in $G$.

Each connection $\omega$ on $P$ gives rise to a trivialization
$F: {\cal P}_{u_0}P\to {\cal P}_{\pi(u_0)}\Sigma \times C_0^\infty
(I,G)$ defined by $F(\gamma_P)=(\hat\pi (\gamma_P),a)$, where $a$ is
the unique element of $C_0^\infty(I,G)$ such that
$\gamma_P=s_\omega(\hat \pi (\gamma_P))\cdot a$, with $s_\omega$
denoting the unique $\omega$-horizontal lift of a path in ${\cal
P}_{\pi(u_0)}\Sigma$ to ${\cal P}_{u_0}P$. Explicitly, we have

$$
a(t)={\rm P}\, \exp \left(\int_0^t \omega(\dot\gamma_P(s))
ds\right).\eqno(7.9)
$$

\noindent This path-ordered exponential of the connection form
along arbitrary paths in $P$ (not just along the horizontal paths
in $P$ usually considered) is an important ingredient in Fulp's
work.

Then two types of connections on path space are introduced.
Firstly, any connection $\omega: TP\to${\bf g} on $P$ induces a
connection $\hat\omega:T({\cal P}P)|_{\gamma_P} \to
C^\infty(I,${\bf g}) ``pointwise".
Secondly, one can use a construction analogous to that of the
canonical flat connection on the trivial bundle $P(\Sigma,G)$ (namely
as the pullback to $P$ of the Maurer-Cartan form $\theta$ on the
fibre, with respect to the fibre projection $P\to G$ [KobNom]) to
obtain flat connections on loop space.
The relation with Polyakov's work [Pol2], who also
introduced flat connections on path space, remains somewhat
elusive, which indicates that it may be hard to make it
well-defined in the rigorous mathematical framework presented
here. Also the eventual use of the path space ${\cal P}P$
in gauge theory may be limited, since its bundle structure is not
compatible with parametrization invariance.

Also the paper by Gross [Gro] aims at providing a mathematical framework for
some gauge theoretical results obtained
by physicists. He chooses to work with the
space ${\cal P}\R^n$ of piecewise smooth paths in $R^n$, based at the origin
(the generalization to general manifolds of some of the results is
discussed briefly).
This path space is a normed vector space with respect to the norm

$$
||\g||=\int_0^1 |\dot\g (t)|\,dt \eqno(7.10)
$$

\ni Next, path forms are introduced, which differ from the ones discussed
above in that they are $k$-forms on $\R^n$ attached to the endpoints $\g(1)$ of
paths $\g$ from ${\cal P}\R^n$. Gross' central point of interest are path
two-forms $h$ with values in the Lie algebra {\bf g}, i.e. for two
vectors $\vec u$ and $\vec v$ in $\R^n$, and $\g\in{\cal P}\R^n$, $h(\g)<\vec
u,\vec v>\in ${\bf g}. A particular case of such two-forms are the
so-called lasso forms $L(\g)$ associated with $C^\infty$ {\bf g}-valued
one-forms $A$ on $\R^n$ according to

$$
L(\g)<\vec u,\vec v>=U_\g^{-1}\,F(\g(1))<\vec u,\vec v>\,U_\g \eqno(7.11)
$$

\ni where $F$ is the field strength of $A$. It is argued that these are good
variables for Yang-Mills theory, and it is therefore important to characterize
the non-linear embedding of the subspace of lasso forms in the space of all
smooth path two-forms. Necessary and sufficient conditions are given under
which a general $h(\g)$ is a lasso form. Furthermore, Gross derives a
generalization of Stokes' theorem, i.e. a non-abelian analogue of $\int_{\del
S}A=\int_S F$ in terms of path variables (other references on the non-abelian
Stokes' theorem include [Are2, Bra, FiGaKa]). Yang-Mills equations on path
space first obtained by Bialynicki-Birula, Mandelstam and others (cf. Sec.11
below) are rederived and discussed. In fact, the lasso forms (7.11) are closely
related with the variables (11.3). Gross' paper contains many useful references
to both the mathematics and physics literature on the subject.

\vskip1.5cm

\line{\ch 8. Identities satisfied by the (traced) holonomies\hfil}

We now give some more properties of the untraced and traced holonomies,
which follow from their definitions (3.4) and (4.2). Take $G$ to be the
gauge group $GL(N,\C)$ or one of its subgroups ($U(N),\,SU(N),\,SO(N),$
etc.) in its fundamental representation in terms of $N\times N$ (complex)
matrices. For two loops $\g_1$ and $\g_2$ based at $x$, we have

$$
U_{\g_1\circ\g_2}(A)=U_{\g_2}(A)\cdot U_{\g_1}(A),\quad\forall A,\eqno(8.1)
$$

\ni with the loop product as defined in (7.1), the dot denoting matrix
multiplication and $U_{\g}$ defined by (4.1). In other words, the
mapping $U$ is compatible with the product structure on loop space. For
the inverse loop $\g^{-1}$ of $\g$, (7.2), we have

$$
U_{\g^{-1}}=(U_\g)^{-1}.\eqno(8.2)
$$

\ni Here and in the rest of this section, dependence of $U_\g$ and
$T(\g)$ on $A$ is understood.
Since (8.2) holds also for open paths $\g$, we derive the important
{\it retracing identity}

$$
U_\g=U_{\g'},\quad{\rm for }\; \g'=((\g_1\circ w)\circ w^{-1})\circ\g_2,
\eqno(8.3)
$$

\ni where $\g_1\circ\g_2=\g$, $w$ is an open path ``glued to $\g$"
(see Fig.3) and we have generalized the composition law (7.1) to open paths.

\epsffile{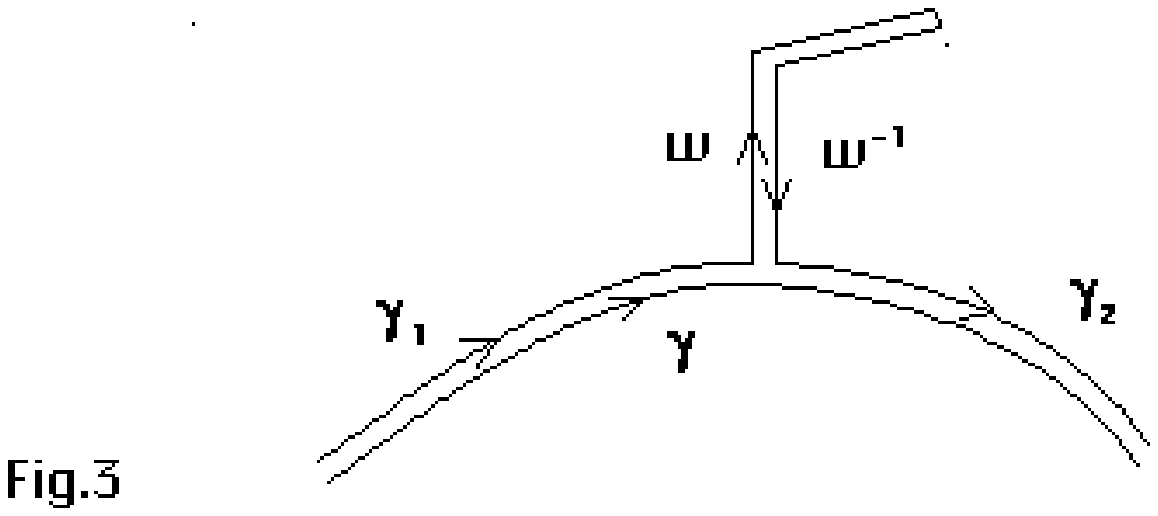}

Although the composition of loops is non-associative,
it is easily seen that $U_{\g'}$ in (8.3) is independent of
the order of composition of the loop segments; for
example, we could have chosen $\g'=\g_1\circ (w\circ (w^{-1}\circ\g_2))$
instead. The property (8.3) motivates the introduction of an equivalence
relation on paths. A loop $\g$ is called a {\it thin loop} if it is homotopic
to the trivial point loop $p$ by a homotopy of loops whose image lies
entirely within the image of $\g$. Two loops (paths) $\g_1$ and $\g_2$ are
called thinly equivalent if $\g_1\circ\g_2^{-1}$ is a thin loop. Note that
two loops that differ by an orientation-preserving reparametrization are
always thinly equivalent.

We will denote the equivalence class of a loop $\g$ by $\bar\g$.
Under the above equivalence relation, the set
of equivalence classes $\Omega_{x_0}\Sigma/\sim$ acquires a group structure,
which is characterized by the relations

$$
\eqalign{&\bar\b=\bar\g\qquad\qquad
      \hbox{if $\b=\g$ mod retracing/reparametrization}\cr
         &\bar\g\circ\bar\g^{-1}=\bar p_{x_0} \cr
         &(\bar\a\circ\bar\b)\circ\bar\g=\bar\a\circ (\bar\b\circ\bar\g)
      \qquad\hbox{(associativity)}\cr}\eqno(8.4)
$$

\ni with the unit $\bar p_{x_0}$, the equivalence class of the constant loop
(2.3). This group is usually called the ``loop group" or the ``group of loops"
(not to be confused with the loop groups of maps from $S^1$ into a Lie
group $G$ \`a la Pressley and Segal [PreSeg]), and will be denoted by ${\cal
L}\Sigma$. The group ${\cal L}\Sigma$ is non-abelian and, as shown by
Durhuus and Leinaas [DurLei],
not locally compact (in the ``area topology" induced by (6.3) in Sec.6), which
means it is ``very large" even locally. This feature leads to complications
when one tries to define a Fourier transform on ${\cal L}\Sigma$. The loop
group has been discussed by many authors in view of its application to physics
[Ana1, AshIsh1, Bar, DurLei, Fis, Fr\"o, GamTri3, Men1]. Its construction
has been around in the mathematics literature for a
long time (see the remarks and references in [Bar]). Some toy examples of
simpler loop groups in two dimensions are contained in [DurLei].
A simplified version for polygonal loops on $\R^4$ is discussed in [GamTri3].
- Some authors prefer to induce an equivalence relation on loops by defining
$\b\sim\g$ if $U_\b(A)=U_\g(A)$, $\forall A$, which explicitly refers to
the space of connections [Lew, AshLew2]. Although sometimes assumed, there
is as yet no proof that this definition is equivalent to the thin equivalence
introduced above.

The infinitesimal
generators of the group ${\cal L}\Sigma$ are the ``lasso forms"
$L(\g)<\vec u,\vec v>$ already introduced in (7.11),
where $\g$ is a path originating at the base point $x_0$,
and $\vec u$ and $\vec v$ are two tangent vectors in
the point $\g (1)$. More precisely, the linear span of all objects $L(\g)<\vec
u,\vec v>$ with fixed base point $x_0$ is the Lie algebra of the restricted
holonomy group at $x_0$ [AmbSin, Gro, Tel]. These objects have also been used
in the work by Gambini and Trias [GamTri3,4].

Note that by virtue of (8.1), the
holonomy mapping $U$ may be viewed as a group homomorphism $U:{\cal
L}\Sigma\rightarrow G$. The Wilson loops (4.2) may therefore be regarded as
the characters of a representation of the loop group ${\cal L}\Sigma$ [Ana1,
DurLei, Fr\"o].

There have been several suggestions of embedding the group ${\cal L}\Sigma$
into certain large groups with ``nice" properties. Di Bartolo et al.
introduce a set of distributional functions $X(\g)$ on loop space,
so-called multi-tangents, defined by the
decomposition

$$
U_\g(A)=\one +\sum_{n=1}^\infty\int d^3x_1\dots d^3x_n\;A_{a_1}(x_1)\dots
A_{a_n}(x_n)\,X^{a_1 x_1\,a_2 x_2\dots a_n x_n}(\g)\eqno(8.5)
$$

\ni [DiGaGr]. Then each loop $\g$ gives rise to a set of $X(\g)$ satisfying
certain algebraic and differential properties. However, there exist more
general objects $X$ which satisfy the same properties but do not correspond
to any loop $\g$ in ${\cal L}\Sigma$. They are identified with elements of
an {\it extended loop group}. This extended group contains elements $\g^q$
(the loop $\g$ traversed $q$ times), where $q$ may be any real number.
Tavares [Tav] works with iterated Chen integrals of one-forms $\omega_i$,

$$
\int_\g \omega_1\dots\omega_r:=\int_0^1 dt_1\int_0^{t_1}dt_2\dots\Big(
\int_0^{t_{r-1}}dt_r\,\omega_r(t_r)\Big)\dots\omega_1(t_1).\eqno(8.6)
$$

The set of all such objects gives rise to the so-called shuffle algebra
$Sh(\Sigma)$. A generalized loop is now taken to be a continuous complex
algebra homomorphism $Sh(\Sigma)\rightarrow\C$ that vanishes on a certain
ideal in $Sh(\Sigma)$. The set of all generalized loops can be made into a
group of which ${\cal L}\Sigma$ is a subgroup. Note that the holonomy $U_\g
(A)$ is a special element of $Sh(\Sigma)$ (c.f. expression (3.4)), namely
an infinite sum of
iterated integrals, with $\omega_1=\omega_2=\dots=A$, with $A$ a {\bf g}-valued
one-form. In general, iterated integrals do not have a simple behaviour
under gauge transformations.

There are analogous identities satisfied by the traced holonomies,
characterizing them as a particular subset of complex-valued functions
on $\Omega\Sigma\times {\cal A}$ (more precisely, $\Omega\Sigma\times {\cal A}/
{\cal G}$ because of their gauge invariance).
Independently of the gauge group $G$, we have the identities

$$
T_A(\g_1\circ\g_2)=T_A(\g_2\circ\g_1)\eqno(8.7)
$$

\ni because of the cyclicity of the trace, and again a retracing identity,

$$
T_A(\g)=T_A(\g'),\quad \g,\g' \hbox{ related as in (8.3)}.\eqno(8.8)
$$

\ni Another set of identities are the so-called Mandelstam constraints,
whose form depends on the dimension $N$ of the group matrices. They can
be systematically derived from the identity of $N$-dimensional
$\d$-functions,

$$
\sum_{\pi\in {\cal S}_{N+1}}(-1)^{\sigma(\pi )}\, \d_{i_1,\pi (j_1)}\dots
      \d_{i_{N+1},\pi (j_{N+1})}=0,\quad i_k,j_k=1\dots N,\eqno(8.9)
$$

\ni with the sum running over all permutations $\pi$ of the symmetric
group of order $N+1$, ${\cal S}_{N+1}$, and $\sigma (\pi)$ denoting the
parity of the permutation. Contracting $N+1$ holonomy matrices $U_\g$
with (8.9) results in a trace identity for (combinations of) $N+1$ loops.
For $N=1$, we have

$$
T_A(\a) T_A(\b)-T_A(\a\circ\b)=0,\eqno(8.10)
$$

\ni and for $N=2$,

$$
T(\a) T(\b) T(\g)-\frac12\Big( T(\a\b) T(\g) +T(\b\g)T(\a)
+T(\a\g)T(\b)\Big)+\frac14\Big(T(\a\b\g) +T(\a\g\b)\Big)=0,\eqno(8.11)
$$

\ni etc., where we have omitted the subscript $A$ and the symbol $\circ$
denoting loop composition.
Note that the Mandelstam identities are non-linear algebraic
equations on the functions $T$. If we want to consider traced holonomies
of specific subgroups of $GL(N,\C)$, there will be more identities
satisfied by $T$, for example, deriving from a condition $\det U_\g=1$
(see [GliVir, GamTri6] and the next section for some selected cases).
Berenstein and Urrutia [BerUrr] discuss the derivation of Mandelstam
identities from the characteristic polynomials of matrices and extend this
to the case of supermatrices.

We may use the functions $T$ to induce an equivalence relation on
the loop space $\Omega\Sigma$ by defining

$$
\b\sim\g\qquad {\rm if}\qquad T_A(\b)=T_A(\g),\;\forall A.\eqno(8.12)
$$

\ni The composition law for equivalence classes $\bar\g$,

$$
\bar\g_1\circ\bar\g_2:=\overline{\g_1\circ\g_2},\eqno(8.13)
$$

\ni induces an abelian group structure on $\Omega\Sigma/\sim$ {\it if} in
addition the relation $T_A(\a)=T_A(\b)\Rightarrow
T_A(\a\circ\g)=T_A(\b\circ\g)$ is satisfied for all $\g$ [GamTri6, AshIsh1].

\vskip1.5cm

\line{\ch 9. Equivalence between gauge potentials and holonomies\hfil}

The importance of the (traced) holonomies lies in the fact that from them one
can reconstruct gauge-invariant information about the gauge potential $A$.
The contents of the so-called reconstruction theorems is to specify a set of
algebraic and differential conditions on a set of functions on loop space
which ensure that from them one can uniquely compute the corresponding
equivalence class $[A]\in {\cal A}/\cal G$.

The best-known case is that of holonomies based at a point, for which various
versions of the reconstruction theorem are available, depending on the
mathematical setting. Although known to mathematicians for a long time, they
have been regularly rediscovered by physicists. A very detailed discussion and
derivation is contained in the paper by Barrett [Bar]. I quote here his
reconstruction theorem for differentiable principal fibre bundles (related
treatments may be found in [Ana1, Fr\"o, Gil, GliVir]):

\ni {\it Reconstruction Theorem.} Suppose $\Sigma$ is a connected manifold
with basepoint $x_0$ and the map $H: \Omega_{x_0}\Sigma\rightarrow G$
satisfies the following conditions:

\item{(i)} $H$ is a homomorphism of the composition law of loops,
$H(\g_1\circ\g_2)=H(\g_2)H(\g_1)$.
\item{(ii)} $H$ takes the same values on thinly equivalent loops:
$\g_1\sim\g_2$ if $\g_1\circ\g_2^{-1}$ is thin (cf. Sec.8).
\item{(iii)} For any smooth finite-dimensional family of loops
$\tilde\psi:U\rightarrow\Omega_{x_0}\Sigma$, the composite map $H\psi:
U\rightarrow\Omega_{x_0}\Sigma\rightarrow G$ is smooth.

\ni Then there exists a differentiable principal fibre bundle
$P(\Sigma,G,\pi)$,
a point $p\in\pi^{-1}(x_0)$ and a connection $\Gamma$ on $P$ such that $H$ is
the holonomy mapping of $(P,\Gamma,x_0)$. (In (iii) above, $U$ is an open
subset of $\R^n$ (parametrizing the family), for any $n$, and $\tilde\psi$ is
smooth in the sense that the associated map $\psi:U\times I\rightarrow \Sigma$
is continuous and piecewise $C^\infty$ with respect to the the loop parameter
$t\in I$.)

Note that in this formulation the principal bundle $P$ is not fixed a priori,
but only the base space $\Sigma$ and the fibre $G$ are. Fixing $P$ would
amount to fixing the homotopy class of the holonomy mapping. The idea of
simultaneously considering all possible principal bundles for given
$(\Sigma,G)$
is taken up in [Bar] and [Fis], and the space of PFB isomorphism equivalence
classes of triplets $(\Sigma,G,\Gamma)$ is called the ``grand superspace" $\cal
S$ by Fischer. Recalling the definition of the loop group ${\cal L}\Sigma$
introduced in the previous section, one can set up a natural bijection
between $\cal S$ and the space of homomorphisms Hom$({\cal L}\Sigma,G)/G$ (the
quotient by $G$ takes care of the residual gauge freedom of the holonomies at
the base point $x_0$). Fischer observes that one can generalize this
construction by taking the quotient of ${\cal L}\Sigma$ with respect to the
normal subgroup $\cal H$ of loops homotopic to the trivial loop $p_{x_0}$, so
that $\pi_1(\Sigma,x_0)={\cal L}\Sigma/\cal H$. One then obtains a one-to-one
correspondence between the homomorphisms Hom$(\pi_1(\Sigma,x_0),G)/G$ and
triplets $(\Sigma,G,\Gamma)$ where $\Gamma$ is now a {\it flat} connection.
He suggests to look for some other normal subgroup $\cal N$ of ${\cal
L}\Sigma$ such that elements of Hom$({\cal L}\Sigma/{\cal N},G)/G$ correspond
to solutions to the Yang-Mills equations [Fis]. Related ideas are elaborated
on by Lewandowski [Lew], who calls $G_{\cal N}={\cal L}\Sigma/{\cal N}$ the
generalized gauge group associated with ${\cal N}$ and the projection map
${\cal L}\Sigma\rightarrow {\cal L}\Sigma/{\cal N}$ a generalized holonomy
map.

Let us go back to the discussion of the reconstruction theorem, now for the
case of the traced holonomies. Since we already have a reconstruction theorem
for the holonomies, it suffices to show one can recover the holonomies,
(4.1), from the Wilson loops (4.2). The main task here is to find algebraic
conditions analogous to (i) and (ii) above, characterizing uniquely the traced
holonomies as a subset of complex-valued functions on $\Omega\Sigma$. It seems
to be much harder to come up with a set of necessary and sufficient conditions,
moreover, these conditions now depend on the gauge group $G$. To my knowledge,
the problem of giving a complete set of such conditions for a
general gauge group $G$ has not been solved. The most advanced results in
this context are those obtained by Giles [Gil].

The contents of his reconstruction theorem is essentially as follows: Given any
complex-valued function $F(\g)$ on the loop space $\Omega\Sigma$ satisfying the
Mandelstam identities of order $N$ (and possibly some additional identities,
characterizing a specific subgroup of $GL(N,\C)$), retracing and
reparametrization invariance, equation (8.8) and appropriate smoothness
conditions, one can construct (modulo a residual gauge freedom) $N\times N$
matrices $U_\g\in GL(N,\C)$ (or of the subgroup in question) such that the
traces of products $U_{\g_j}\cdot U_{\g_i}\cdot \dots$ are exactly given by
$F(\dots\circ\g_i\circ\g_j)$.

Giles gives an explicit way of reconstructing the holonomies from the traced
holonomies, which is very useful in practical applications. However, his
results
are incomplete in at least two aspects. Firstly, in addition to the usual
Mandelstam constraints, there are inequalities restricting the range of the
Wilson loops, i.e. given a set of $n$ Wilson loops, they can in general not
attain arbitrary complex values, even if all the Mandelstam identities are
fulfilled [Lol5]. Secondly, it has not been proven that by running
through all admissible $F(\g)$'s one recovers indeed {\it all} possible
holonomy
configurations $U_\g$. In general the $U_\g$'s so obtained will form a subgroup
of $G$. This however can only happen if $G$ is non-compact, a well-studied
example being that of $G=SL(2,\C)$ [GoLeSt]. This paper explores the
degeneracies of the Wilson loops and the momentum loop variables (14.2), as
well as their interplay with the dynamics of general relativity.

For the sake of illustration, and because of their importance in many
applications, here are the explicit Mandelstam constraints for $G=SL(2,\C)$, in
the fundamental representation by complex $2\times 2$-matrices. One finds

$$
\eqalign{
(a)\qquad & T(\hbox{point loop } p)=1\cr
(b)\qquad & T(\g_1)=T(\g_2),\qquad \hbox{if $\g_1$ and $\g_2$ are thinly
   equivalent}\cr
(c)\qquad & T(\g_1\circ\g_2)=T(\g_2\circ\g_1)\cr
(d)\qquad & T(\g)=T(\g^{-1})\cr
(e)\qquad & T(\g_1)T(\g_2)=\frac12\Big( T(\g_1\circ\l\circ\g_2\circ\l^{-1})+
            T(\g_1\circ\l\circ\g_2^{-1}\circ\l^{-1})\Big).
}\eqno(9.1)
$$

\ni In (e), $\l$ is a path connecting a point on $\g_1$ with a point on
$\g_2$, as illustrated by Fig.4. Note that (e) implies both (a) and (d).
However, the conditions have been separated in this way since (a)-(c) hold
for any group, whereas (d) and (e) are true just for $SL(2,\C)$. Presumably
the set (9.1) exhaust all the Mandelstam constraints for this particular group
and representation, although I am not aware of the existence of a formal proof.
In any case, it can be shown that there are holonomies $U_\g$ that cannot
be reconstructed from the Wilson loops, namely those that lie in the subgroup
of so-called null rotations. However, in a sense this incompleteness is
negligible in the physical applications considered so far [GoLeSt],
and in fact the Wilson loops are ``as complete as they could be" [AshLew1].

\epsffile{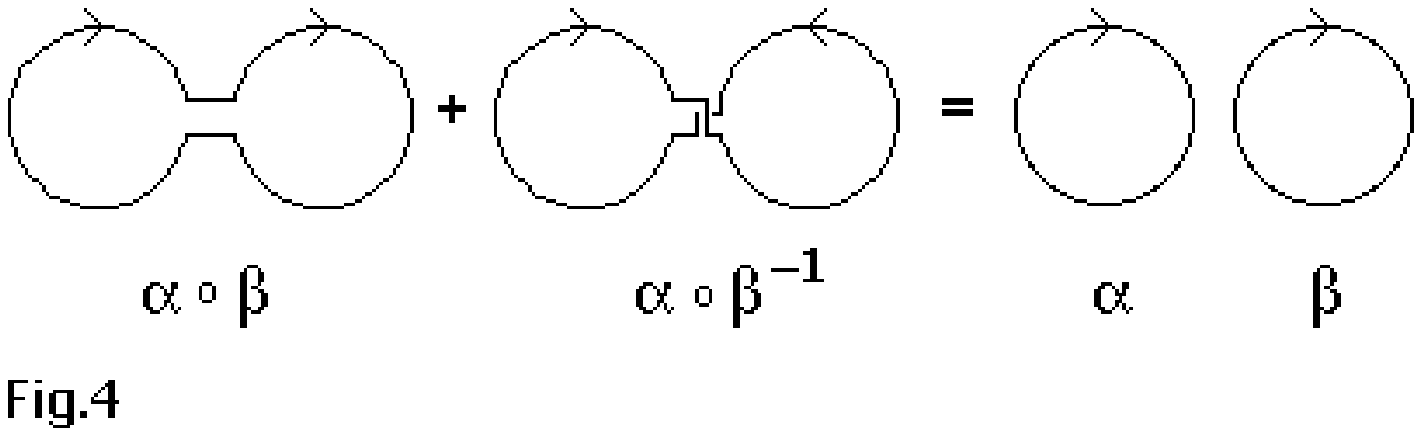}

If one wants to restrict the group $G$ to be described by the Wilson loops to
a subgroup of $SL(2,\C)$, there are further conditions that have to be
imposed on the loop variables, including inequalities. For the subgroup
$SU(2)\subset SL(2,\C)$, one has

$$
\eqalign{
(a)\qquad&T(\g)\hbox{ real}\cr
(b)\qquad&-1\leq T(\g)\leq 1\cr
(c)\qquad&\Big( T(\g_1\circ\g_2)-T(\g_1\circ\g_2^{-1})\Big)^2\leq 4\Big(
1-T(\g_1)^2
  \Big) \Big(1-T(\g_2)^2\Big)\cr
\qquad&\qquad\vdots
}\eqno(9.2)
$$

\ni whereas the analogous conditions for the subgroup $SU(1,1)\subset
SL(2,\C)$ are

$$
\eqalign{
(a)\qquad&T(\g)\hbox{ real}\cr
(b)\qquad&T(\g_1)^2\leq 1\hbox{ and } T(\g_2)^2\leq 1\Rightarrow\cr
\qquad&\Big( T(\g_1\circ\g_2)-T(\g_1\circ\g_2^{-1})\Big)^2\geq 4\Big(
1-T(\g_1)^2 \Big) \Big(1-T(\g_2)^2\Big)\cr
\qquad&\qquad\vdots
}\eqno(9.3)
$$

\ni As indicated by the dots, there are other inequalities for more complicated
loop configurations, involving more than two loops $\g_i$.
Note that all of the above relations hold
independently of $A$, and hence may be interpreted as {\it constraints} on
loop space functions on $\Omega\Sigma$,
rather than as {\it identities} on the loop-dependent functions on the space
of connections, i.e. as functions on $\Omega\Sigma\times {\cal A}/{\cal G}$.
Also at least
some of these algebraic conditions can be interpreted as conditions on the
underlying loop space. For example, (9.1.d) may be interpreted as a
condition that $T$ is a function of unoriented loops.

If one wants to
reformulate a theory purely in terms of loop variables, without making any
reference to the connection variables $A$, one has to come up with a complete
set of conditions of the type discussed above to ensure equivalence with the
usual local formulation. This is a non-trivial task, and not always appreciated
in physical applications. The difficulty of proving rigorous reconstruction
theorems for the Wilson loops has somewhat hampered this kind of ``pure loop
approach". The most progress in isolating the true degrees of freedom in
such an approach has been made on the hypercubic lattice [Lol2,3,4].
-  Once a reconstruction theorem has been proven, the method is potentially
very
powerful, since any gauge-invariant statement in terms of the connection $A$ is
in principle expressible in terms of the traced holonomies $T(\g)$.

Alternatively, one may follow a less radical approach which still incorporates
the gauge potentials. For example, one may define a ($G$-dependent) equivalence
relation on loops by identifying two loops if their associated Wilson loops
agree for all possible configurations $A(x)$ (8.12). The
quotient space of loop space one obtains in this way is of course exactly the
one one would like to use as a domain space when
constructing a formulation exclusively in terms of loop variables.
Another possibility is to postulate the existence of a ``loop transform" in
the quantum theory (cf. Sec.15) which allows one to translate wave functions
and
operators from the connection to the loop representation. This automatically
takes care of algebraic constraints among the Wilson loop operators, i.e.
quantum analogues of the Mandelstam constraints.

Incidentally, there have also been attempts to reformulate gauge theories in
terms of the field strengths $F$ instead of the gauge potentials $A$, because
they have a simpler, homogeneous transformation behaviour. However, it is well
known that gauge-inequivalent connection configurations may have the same
field strength (although this does not happen for ``generic" connections), and
hence a rigorous reconstruction theorem does not exist. A discussion of this
degeneracy and a guide to the literature is contained in the paper by Mostow
and Shnider [MosShn].

\vskip1.5cm
\vfill\eject 

\line{\ch 10. Physical interpretation of the holonomy\hfil}

It is well known that the existence of a connection on a manifold enables one
to define a notion of {\it parallel transport}. For a field $\Psi(x)$
transforming according to the defining representation of the gauge group
$G$,

$$
\Psi(x){\buildrel g\over\longmapsto} g^{-1}(x)\Psi(x),\eqno(10.1)
$$

\ni we can compare fields at different points $x$ and $y$ by
parallel-transporting $\Psi(y)$ from $y$ back to $x$ using the holonomy
$U_w(s_1,s_0)$ along a path $w$ with $w(s_1)=x$, $w(s_0)=y$, to obtain

$$
\Psi_{par,w}(x):=U_w(s_1,s_0)\Psi(y),\eqno(10.2)
$$

\ni which has now the same behaviour under gauge transformations as
$\Psi(x)$. In the limit as the length of $w$ goes to zero, this procedure
leads to the definition of the covariant derivative of $\Psi$ [Pol3].

The holonomy of small closed loops measures the {\it curvature (or field
strength) in internal space}. For a small square loop $\g$ of side
length $\e$ in a coordinate chart around the point $x\in \Sigma$, the
base point of $\g$, the holonomy $U_\g$ can be expanded as

$$
U_\g=\one + g\,F_{\mu\nu}^a(x)X_a \;\e^2 +O(\e^3),\eqno(10.3)
$$

\ni where $\g$ is defined by its four corners, $(x,x+\e\vec e_\mu ,x+\e
\vec e_\mu
+\e \vec e_\nu, x+\e \vec e_\nu)$, with $\vec e_\mu$ denoting the unit
vector in $\mu$-direction. Note that in the non-abelian gauge theory,
$F_{\mu\nu}(x)=F_{\mu\nu}^a (x) X_a$ is not an observable, since it
transforms non-trivially under gauge transformations.
The idea of measuring the eigenvalues of holonomy matrices by non-abelian
analogues of interference experiments \`a la Aharonov and Bohm is discussed by
Anandan [Ana2,3] (an earlier treatment can be found in [WuYan]). He also
points out differences between the applications to gauge theory and gravity
(albeit not in the Ashtekar formulation) [Ana3]. A beautiful geometric
description of the modified parallel transport associated with the Ashtekar
connection in gravity is given in the review paper by Kucha\v r [Kuc].

Performing a similar expansion for the traced holonomy of the infinitesimal
loop $\g$, we obtain

$$
T_A(\g)={\rm Tr}\,U_\g=1+\frac{g}{N}\sum_a F_{\mu\nu}^a(x){\rm Tr} X_a \,\e^2+
\frac{g^2}{N}\sum_{a,b}
F_{\mu\nu}^a(x)F_{\mu\nu}^b(x)\,{\rm Tr} X_a X_b\;\e^4+O(\e^5),\eqno(10.4)
$$

\ni (no sum over $\mu,\nu$). For a semi-simple Lie algebra {\bf g} we can
always find a basis of generators $X_a$ such that ${\rm Tr}\, X_a X_b=\d_{ab}$.
Moreover, for $G=SU(N)$ (more generally, for any subgroup of the special linear
group $SL(N,\C)$, we have ${\rm Tr}\, X_a=0$, and
the expansion therefore reduces to

$$
T_A(\g)=1+ \frac{g^2}{N}\sum_a F_{\mu\nu}^a(x) F_{\mu\nu}^a(x)\; \e^4 +O(\e^5).
\eqno(10.5)
$$

\ni These expansions illustrate the way local gauge-invariant
information about the curvature or field strength $F$ is contained in
$T_A(\g)$.

Having described the basic mathematical properties of the holonomy and the
traced
holonomy, there remains the question of the physical interpretation of
the underlying loop space itself, and therefore of the way in which a
physical theory is to be constructed on it. Formulating a theory in terms of
non-local variables depending on loops is potentially very different from
the usual local formulations. It makes a difference whether one wants to
use loops as convenient
auxiliary labels in an otherwise local formulation, or postulates
them to be the basic entities of a new, intrinsically non-local
description of gauge theory or gravity. There is a variety of ways in which
past
research has made use of path- and loop-dependent quantities, some
examples of which will be described below. It is important to realize that
different a-priori physical interpretations of the role of loops
suggest different mathematical structures for their
description, for example, the initial choice of a loop space or a
quotient of a loop space. If one thinks of paths or loops as describing
actual trajectories of charged particles [Wil, Bar, Pol3], say,
one may work with smooth $C^\infty$-loops in a (semi-)classical description
and nowhere differentiable loops (which are supposed to give
the main contribution to the Feynman path integral) in the quantum theory.

Many conceptual issues still remain unanswered.
It is not clear whether a non-local description in terms of
holonomies is necessarily tied to the {\it quantum}
aspects of the theory. Also it is clearly the non-linearity of Yang-Mills
theory
and gravity that  calls for a non-local loop formulation,
since the abelian U(1)-theory does not require such a description, although it
is possible [Ash1, AshRov, GamTri2]. One may also ask how big loops are,
and whether
they have a preferred size in a given theory (which could serve to provide a
new
fundamental length scale).
We are lacking a physical argument to decide whether the (traced)
holonomies should be taken
as the basic variables (cf. Polyakov's ``rings of glue" [Pol2]), or
whether genuine physical observables are
again {\it composites} of the elementary Wilson
loops. - The following sections deal with some issues peculiar to loop
formulations of Yang-Mills theory and gravity. They are of an introductory
nature, and to be regarded as a commented guide to the literature.

\vskip1.5cm

\line{\ch 11. Classical loop equations\hfil}

In this section I give some examples of path-dependent formulations
of gauge theory, starting from Mandelstam's early treatment of
electrodynamics coupled to a scalar field, and its subsequent
generalization to the non-abelian case. This illustrates how classical
equations of motion for path-dependent variables may be obtained.
Unfortunately, no analogous classical loop equations have so far been
derived for general relativity.

Mandelstam's description of ``QED without potentials" [Man1] is
interesting in the present context, because it is the first instance
of the use of holonomy-dependent field variables. He demonstrates
that a gauge-invariant and Lorentz-covariant formulation of
electrodynamics (avoiding the unphysical negative-norm states of the
Gupta-Bleuler approach) is possible, {\it provided} one allows for
non-local matter field variables. Instead of using the gauge
potential $A_\mu (x)$ and the charged scalar field $\phi(x)$, he
works with the field strength $F_{\mu\nu}(x)$ and objects

$$
\Phi(\g,x):=\phi(x) \exp \{-ie\int\limits_{-\infty}^{x} A_\mu
d\g^\mu\}\eqno(11.1)
$$

\ni as the basic variables, where $\g$ denotes a space-like path in
Minkowski space originating at spatial infinity and ending at the
point $x$. These variables are obviously invariant under the gauge
transformations

$$
\Phi\rightarrow e^{ies}\Phi,\qquad A_\mu\rightarrow A_\mu+\del_\mu s,
\eqno(11.2)
$$

\ni where $s$ is an arbitrary scalar field satisfying the boundary
condition $s(\pm\infty)=0$. Mandelstam emphasizes that the need for
a gauge-invariant formulation arises in the quantum theory, because
only commutators between (gauge-invariant) observables are well
defined. He argues that the inherent path dependence of the matter
field variable is natural and corroborated by the Aharonov-Bohm
effect.

Bialynicki-Birula [Bia] extended these ideas to formulate what one might
call ``Yang-Mills theory without field strengths". Since the
non-abelian gauge field interacts with itself, there is in a first
step no need to couple it to an external matter field in order to
make the theory non-trivial. Since the field variables $F_{\mu\nu}(x)$
are themselves not observables (i.e. not gauge-invariant),
Bialynicki-Birula introduces in analogy with (11.1) the gauge-invariant
fields

$$
{\cal F}_{\mu\nu}(\g,x):={\rm P}\,\exp\{-ig\int\limits_{-\infty}^x
A_\mu d\g^\mu\}\; F_{\mu\nu}(x)\; {\rm P}\,\exp\{ig\int\limits
^{-\infty}_x A_\mu d\g^\mu\},\eqno(11.3)
$$

\ni which makes use of the holonomies of the space-like paths
$\g$ and $\g^{-1}$, as illustrated by Fig.5.

\epsffile[0 40 436 179]{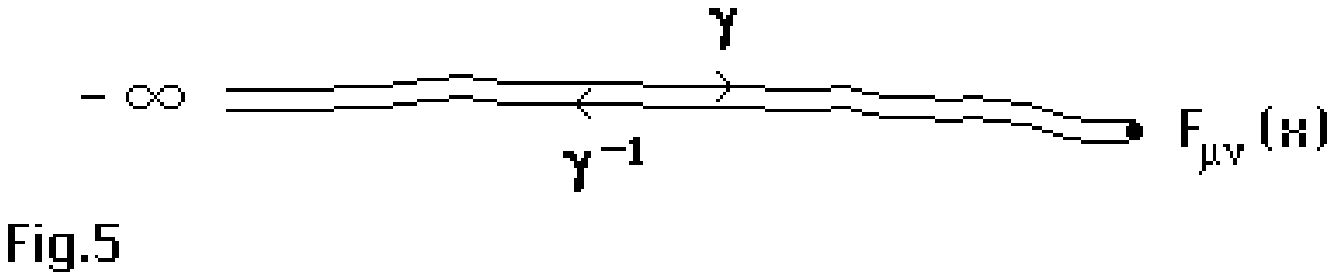}

\ni A corresponding path-dependent potential is introduced and discussed by
Gambini and Trias [GamTri1].
The usual classical Yang-Mills equations of motion

$$
{\cal D}_\mu F^{\mu\nu}_a(x)=(\del_\mu \d_{ac}-g\e_{abc} A_{b\mu}(x))
F^{\mu\nu}_c (x)=0,\eqno(11.4)
$$

\ni with $\e_{abc}$ denoting the structure constants of {\bf g},
translate to

$$
\del_\mu (x) {\cal F}^{\mu\nu}(\g,x)=0,\eqno(11.5)
$$

\ni where the differential $\del_\mu (x)$ acts on the holonomy
$U_{\g,x}$ as an ``endpoint derivative" (see equation (12.1) for a
definition). The
Bianchi identities (which are satisfied automatically in the
connection formulation) have to be imposed as separate equations,

$$
\del_\lambda (x) {\cal F}_{\mu\nu}(\g,x)+\del_\mu (x){\cal F}_{\nu\lambda}
(\g,x)+\del_\nu (x){\cal F}_{\lambda\mu}(\g,x)=0.\eqno(11.6)
$$

\ni It may be somewhat surprising that equations (11.5) and (11.6),
unlike equations (11.4), are {\it linear} in the basic field
variables. Mandelstam [Man2] writes: ``The field equations are simpler
{\it in appearance} than the Maxwell equations of electrodynamics,
since there is no additional current term". He is obviously aware
of the fact that difficulties associated with the non-linear
functional form of the Yang-Mills equations must be hidden
in (11.5), however, neither spelling out in what sense non-triviality
arises, nor making any further use of the equations (11.5)
classically (they are used to obtain equations for the corresponding
path-dependent Green's functions in the quantum theory though).
A rigorous mathematical derivation of the equations (11.5) and (11.6) is
contained in the paper by Gross [Gro], within the mathematical framework
discussed already in Sec.7.

Slightly different gauge-{\it co}variant field variables are used by
Polyakov [Pol1,3], namely,

$$
{\cal F}_\mu (\g,s):= U_{\g}(0,s)\, F_{\nu\mu}(\g(s))\,\dot\g^\nu(s)\,
U_{\g}(s,0),\eqno(11.7)
$$

\ni where $\g$ is a closed loop, $s$ some intermediate parameter
value, $s\in [0,1]$, and $U_{\g}(0,s)$ the holonomy along the
portion $[0,s]$ of the loop $\g$ (Fig.6)

\epsffile[0 640 576 840]{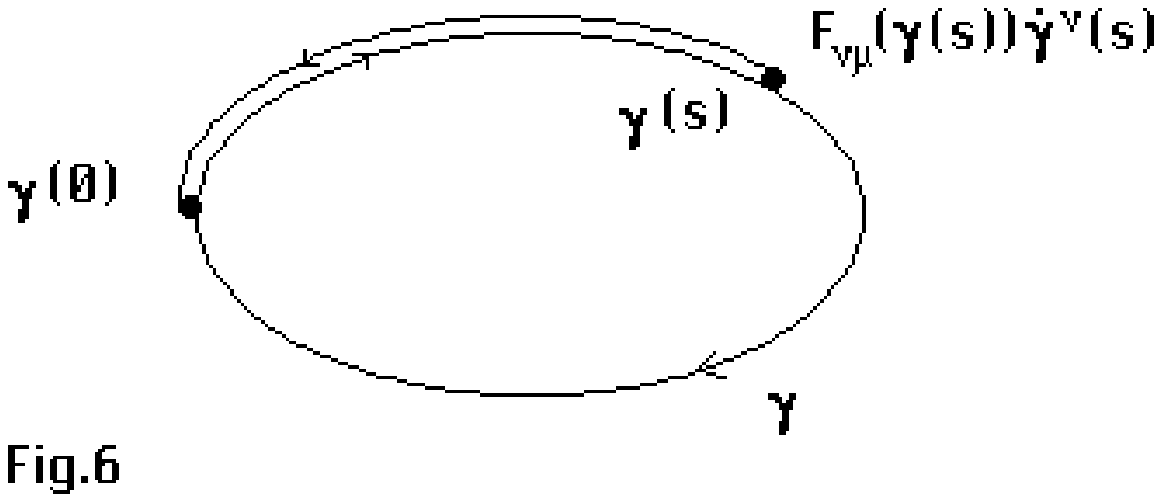}

\ni The field equations

$$
\frac{\d}{\d x^\mu(s)} {\cal F}^\mu (\g,s):=
\dot\g^\lambda(s)\frac{\d}{\d \sigma^{\lambda\mu}(\g(s))} {\cal F}^\mu
 (\g,s)=0\eqno(11.8)
$$

\ni correspond to the Yang-Mills equations projected onto the tangent
vector $\dot\g^\nu$ of $\g$ at $s$. The differential operator
$\frac{\d}{\d \sigma^{\lambda\mu}(\g(s))}$ is the projection of the area
derivative of definition (12.3), obtained by adding an infinitesimal loop in
$\mu$-direction at the point $s$. Note that ${\cal F}_\mu$ still
depends on the curve parameter $s$; parametrization invariance
and the (projected) Bianchi identities have to be imposed as
separate equations in addition to (11.8). The reason why this
reformulation of Yang-Mills theory was thought to be appealing is
its close formal resemblance with the two-dimensional non-linear
$\sigma$-model (see also the work of Aref'eva [Are1,3,4] for a related
treatment). It turns out that this simple analogy does not
go through, essentially because the theory is not defined on
ordinary space, but on loop space. The loop equations (11.8), unlike
their analogues for the classical non-linear $\sigma$-model, do not
contain enough information for constructing a set of conserved
currents [GuWan]. In Polyakov's words, the (up-to-date) failure
of this approach ``has to do with the difficulties we experience
in treating equations in loop space, or, what is more or less the
same, with string dynamics"[Pol3].

A similar framework is investigated by Hong-Mo, Scharbach and Tsun [HoScTs,
HonTsu]. Like Polyakov, they work with parametrized loops and try to exploit
the idea that ${\cal F}_\mu$ (11.7) constitutes a ``connection on loop space".
The Bianchi identity in loop space then amounts to the statement that the
curvature of this connection vanishes, and hence the loop space connection
${\cal F}_\mu$ is flat ([Pol1,2], see also [Ful2]):

$$
\frac{\d}{\d x^\nu(s')} {\cal F}^\mu (\g,s)-
\frac{\d}{\d x^\mu(s)} {\cal F}^\nu (\g,s')+
[{\cal F}_\mu (\g,s),{\cal F}_\nu (\g,s')]=0.
\eqno(11.9)
$$

\ni According to Hong-Mo et al., the non-flatness of the connection
${\cal F}_\mu$, i.e. the non-vanishing of the
right-hand side of (11.9), is related to the presence of a non-abelian
monopole charge. They also derive a reconstruction theorem for the gauge
potential $A_\mu (x)$ from the loop space potential ${\cal F}_\mu$, and
establish an action principle in loop space, using a formal functional
integral over the space of parametrized loops.
In order to avoid the explicit parametrization dependence of equations like
(11.8) and (11.9), Gambini and Trias in a related work
reformulate them in terms of geometric,
but distributional quantities [GamTri4].

A further example of loop-dependent
equations of motion for Yang-Mills theory, this time in terms of the traced
holonomies (4.2), is given by the classical Makeenko-Migdal equations [Mig2]

$$
\del^\mu (x) \frac{\d}{\d \sigma^{\mu\nu}(x)}\, T(\g)=0,\eqno(11.10)
$$

\ni with $\del^\mu (x)$ denoting the ordinary differential on
space-time and $\frac{\d}{\d\sigma^{\mu\nu}(x)}$ Mandelstam's area
derivative (see the next section for a definition). What is puzzling
about this equation is its linearity
in the field variable $T$; again the non-linearities seem to
have disappeared. In fact, there are spurious solutions to
(11.10) [Mig2], which have to be eliminated by other means. This
feature is attributed ``partly to the presence of the Mandelstam
constraints", which have not been taken into account in the
derivation of the equation. Migdal concludes that ``the classical
loop dynamics is quite complicated and implicit, but presumably
it is irrelevant as well as the classical colour dynamics".

A similar non-uniqueness for the derivation of equations of motion
is present in the quantum theory. Various methods have been proposed to
derive an equation for the ``Wilson loop average", i.e. the vacuum
expectation value $<T(\g)>$ of the (space-time) Wilson loop, or that of the
holonomy operator, the most famous of which is the Makeenko-Migdal
equation [MakMig1,2, Mig1, BGNS].
In such derivations, one typically makes
use of identities for first and second order functional path derivatives to
arrive at expressions that are supposed to serve as quantum Yang-Mills
equations. (A question of considerable interest in the late seventies was
the relation of
such path-dependent equations with the equations of motion for the quantized
relativistic string; some references are [CorHas, GamGri, GerNev1,
MakMig1,3, Mig1, Nam]).
However, no solutions to these
equations have been found in three and four space-time dimensions. It has
been shown that the $<T(\g)>$ are multiplicatively renormalizable (for loops
$\g$ with a finite number of cusps and self-intersections)
[DotVer, Are3,4,5, BrNeSa, BGNS],
but no useful
equations have been formulated for the renormalized functional $<T(\g)>_R$.

\vskip1.5cm

\line{\ch 12. Differential operators\hfil}

In order to define equations of motion in a path-dependent approach,
one needs to have a notion of differentiation. The properties
of differentials operators are intimately tied to the space they act
upon, hence path and loop derivatives assume different meanings in
different contexts. Unfortunately those distinctions are often not
clearly stated in the literature, nor are the relevant spaces of
loop functions and the underlying loop spaces.

Recall that for differentiation on some space $X$ to be well defined,
one needs locally at least some topological vector space structure on $X$
[ChoDeW].
If one is lucky, $X$ can be made into a Banach space (i.e. a
complete normed vector space), in which case most of the differential
calculus on $\R^n$ can be generalized in a straightforward way to
$X$. Also, the norm induces a translationally invariant
metric and a natural topology on $X$. For more general topological
vector spaces one may still be able to define differentiation, but
there are in general {\it no inverse and implicit function theorems
and no theorems on the existence and uniqueness of solutions of
differential equations}.

In physical applications, $X$ is usually some space of loop
functions or functionals, i.e. essentially an infinite-dimensional
function space with a vector space structure, but is rarely given
any further structure, for example, a topology. A similar statement
concerns the loop and path spaces themselves. In some of the
previous sections we described the problem of ``giving
structure" to these spaces, which in turn is an obstruction to
defining a meaningful differential calculus on them.
There exist several definitions of such operators, which differ in the way
they treat the path parametrization and (often implicitly) by which class
of functions they are supposed to act on. An explicitly
parametrization-invariant geometric loop calculus has been set up and
applied by Gambini and Trias (see, for example, [GamTri3,4, Gam]).

In the following I will describe some typical path-dependent
differential operators, and outline some of the problems associated
with them. For a function $F(w,x)$ depending on an
(unparametrized) path $w$, with $x$
denoting one of its endpoints, we define an ``endpoint derivative"

$$
\del_\mu (x)\, F(w,x):=\lim_{dx_\mu\rightarrow 0}\frac{F(w',x+dx_\mu)-F(w,x)}
{dx_\mu},\eqno(12.1)
$$

\ni where $w'$ is obtained by adding to $w$ an infinitesimal straight
line element $dx_\mu$ in $\mu$-direction. It is used, for example, by
Bialynicki-Birula [Bia] and Mandelstam [Man2,3] for the special case
where $F$ is the holonomy $U_{\g,x}$ of a path starting at spatial
infinity and going to the point $x$. (The endpoint derivative of the holonomy
taken at the endpoint $x$ of an open path is just the gauge potential $A(x)$.)
Gambini and Trias call it ``Mandelstam's covariant derivative" and use it in a
generalized context where $F$ is a $G$-valued function on the set of open paths
modulo reparametrizations and retracing [GamTri3].

Note that it is not necessary to use open paths in order to obtain the gauge
potential $A$ by differentiation of the holonomy. One may as well stick to
loops and use a ``triangular derivative" [Gil, Gu, HoScTs]. Within a
coordinate patch $W$ on the manifold $\Sigma$, choose an origin $x_0$ and
connect each point $x\in W$ in a continuous way to $x_0$ by an open path, i.e.
such that neighbouring points are joined to $x_0$ by neighbouring paths. For
simplicity, we will use the straight lines (with respect to some auxiliary
metric) $\l_{x_0 x}$ starting at the origin and going to $x$ (see Fig.7).

\epsffile[0 610 576 860]{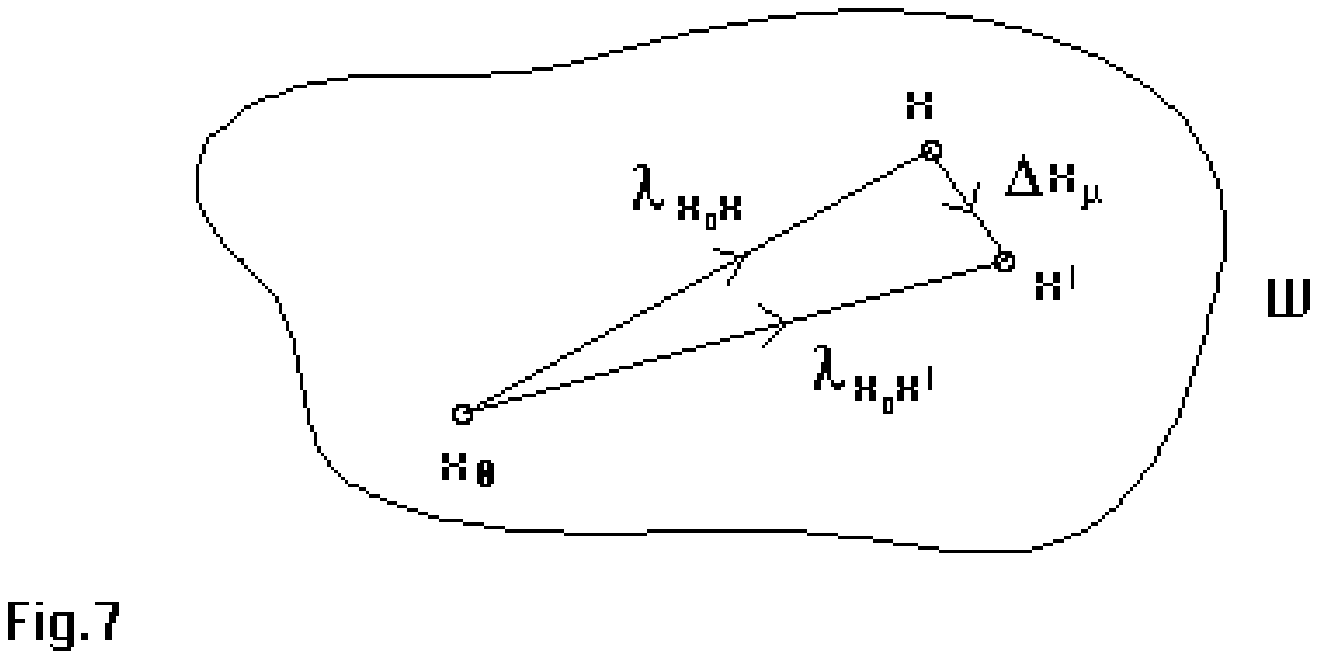}

\ni Consider now a point $x'=x+\Delta x_\mu$ close to $x$ and the closed loop
$\l_{x_0 x'}^{-1}\circ\l_{xx'}\circ\l_{x_0 x}$, where $\l_{xx'}$ is the
infinitesimal straight path linking $x$ and $x'$. The gauge potential may then
be defined as the limit

$$
A_\mu(x)=\lim_{\Delta x_\mu\rightarrow 0}\frac{(U_{\l_{x_0
x'}^{-1}\circ\l_{xx'}
\circ\l_{x_0 x}}-\one)}{||\Delta x_\mu ||}.\eqno(12.2)
$$

\ni This construction is metric-independent. For a different choice
of the paths $\l_{x_0x}$ one obtains a gauge potential that is related to
(12.2) by a gauge transformation. With respect to a fixed metric, the gauge
choice $A_\mu(x) (x^\mu-x_0^\mu)=0$, $A_\mu(x_0)=0$ is sometimes called the
``central gauge" [Gu].

Another frequently used differential operator is the so-called
area derivative. For a path-dependent function $F(w)$ it is
usually defined as

$$
\frac{\d}{\d\sigma^{\mu\nu} (x)} F(w):=\lim_{d\sigma^{\mu\nu}\rightarrow 0}
\frac{F(w\circ_x\g_{\mu\nu})-F(w)}{d\sigma^{\mu\nu}},\eqno(12.3)
$$

\ni where $\g_{\mu\nu}$ is an infinitesimal planar loop in the
$\mu$-$\nu$-plane attached (by path composition) to the  path $w$
at the point $x$ on $w$. The area derivative of a loop function is therefore
a function of loops with a marked point. In a local coordinate chart $\{
x_\mu\}$, the area of the small loop is given by

$$
d\sigma^{\mu\nu}=\frac12 \int_{\g_{\mu\nu}}x^\mu\,dx^\nu,\eqno(12.4)
$$

\ni which is antisymmetric in $\mu$ and $\nu$. For the special case
where $F(w)=U_w$, the holonomy of a path $w$ with initial point
$x_0$ and endpoint $x_1$, we have

$$
\frac{\d}{\d\sigma^{\mu\nu} (x)} U_{w;x_0,x_1}=U_{w;x_0,x}\,F_{\mu\nu}(x)\,
U_{w;x,x_1},\eqno(12.5)
$$

\ni and hence the area derivative is automatically antisymmetric in
$\mu$ and $\nu$. For more general cases of loop functions $F$, one
may have to introduce an explicit antisymmetrization in (12.3) in
order for the definition to make sense. Another derivative that is
sometimes used is obtained by displacing an entire curve $\gamma(s)$ by
an infinitesimal amount $\d\gamma(s)$ (i.e. each point of the loop
$\gamma$ may be shifted) [CorHas, MakMig2]. A way of defining a rigorous
path derivative by introducing a generalized Fourier decomposition for
functions on the unit interval has been described by Gervais and Neveu
[GerNev2].

Further discussions about path and
area derivatives, and their relations to various functional derivatives can be
found in [Ble, BGNS, Br\"uPul, GamGri, Gro, Mig2, Pol2, Tav]. One way of
defining an ``anti-area derivative" (which does not coincide with a simple
surface integration) is discussed by Durhuus and Olesen [DurOle]. For the
application to gravity one wants to avoid the explicit use of the metric in the
definition of those derivatives. Special attention to this demand is given in
[Br\"uPul, Tav].

Let me again emphasize that for
general loop functions there is no reason for the limits in (12.1-3)
to be well defined and exist. Furthermore, if we talk about
``infinitesimal loops", this implies we have chosen some topology on
loop space which tells us about small variations, i.e. what it means
for two loops to be infinitesimally close to each other.
Our ``intuitive" notion of
closeness of two paths is that coming from viewing them as embedded
in the manifold $\Sigma$ and using the Euclidean metric of $\R^n$
in local charts of $\Sigma$. However, this may not be the
appropriate thing to do. For example, in the context of the
``group of loops" introduced in Sec.7 above, two paths are considered
equivalent if they are in the same class with respect to retracing.
This leads to a generalized and non-local (with respect to the
metric topology on $\R^n$) notion of closeness, as has been explained
elsewhere in these notes. Since the issue of
giving topological structures to infinite-dimensional spaces is mathematically
involved, there is a genuine need for physical arguments to restrict
the possible choices, and thus obtain
meaningful equations of motion in a path-dependent approach.

\vskip1.5cm
\line{\ch 13. Lattice gauge theory\hfil}

The only way of obtaining quantitative results about the behaviour
of non-abelian gauge theory, such as the values of hadron masses, and
testing the hypothesis of quark confinement, are in a regularized
version of the theory, where continuous space-time is approximated
by a finite hypercubic lattice. The basic gauge field variables in this
case are the link holonomies $U_l$, i.e. gauge potentials integrated
over elementary lattice links.

\epsffile[0 550 576 860]{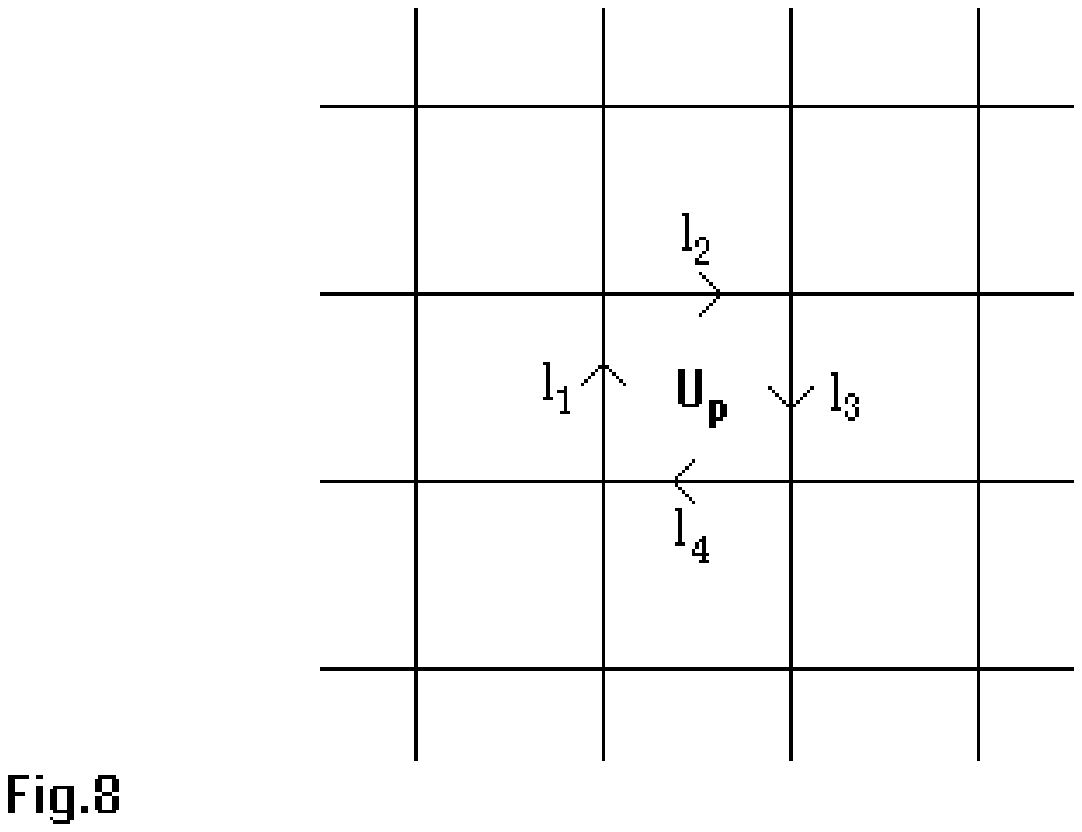}

\ni Wilson, who first proposed this approach to gauge theory [Wil],
identified closed flux lines on the lattice as natural
gauge-invariant objects, and introduced a discretized, Euclidean
form of the Yang-Mills action, which in the continuum limit (as the
length $a$ of lattice links goes to zero) reduces to the ordinary
one. For gauge group SU(N), it is given essentially as the sum over all
lattice plaquettes (elementary square loops made up of four contiguous
oriented links) $P$ of
the traces of the corresponding holonomies $U(P)$,

$$
S_W(U_l)=-\frac{1}{N g^2}\sum_P \,(N-{\rm Tr}\, U_P).\eqno(13.1)
$$

\ni The plaquette holonomy $U_P$ is to be thought of as the
product of the four link holonomies $U_l$ associated with $P$ (see
Fig.8, where we have $U_P=U_{l_1}U_{l_2}U_{l_3}U_{l_4}$).
The underlying physical interpretation for holonomies of closed
paths is that of weight factors in the Feynman path integral,
associated with classical quark trajectories. More precisely, in
order to compute the current-current propagator for quark fields
between two points $0$ and $x$ in space-time, one has to average
over all possible classical quark trajectories and classical gauge
field configurations. The relevant quark configurations are pairs of
quarks created at the origin $0$ and annihilated at $x$ (Fig.9a),

\epsffile[0 640 576 840]{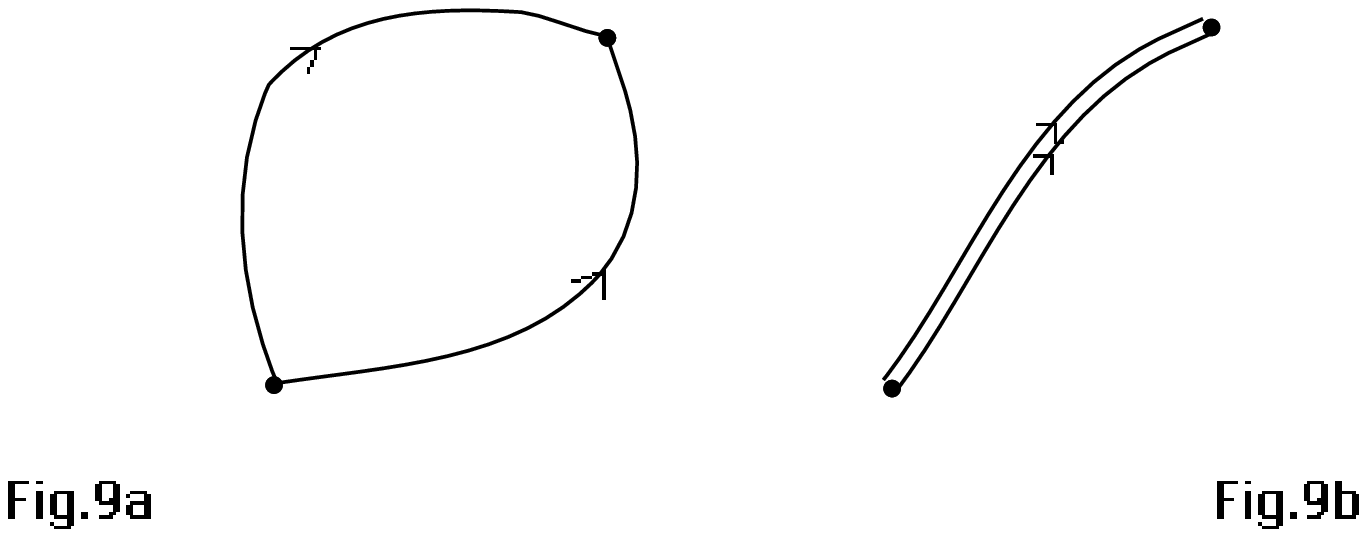}

\ni the
weight for such a pair being exactly the holonomy $U_{\g}$ along the
loop formed by the pair of quark trajectories. In the strong
coupling limit of the lattice gauge theory ($g\rightarrow\infty$),
one can produce arguments that the gauge field average of $U{\g}$
for a fixed lattice loop $\g$ behaves as $\exp icA(\g)$, with
$A(\g)$ denoting the area enclosed by the loop $\g$, and constant $c$.
This suggests confining behaviour for quarks, because large areas
$A(\g)$ (corresponding to the quarks being far apart) are
strongly suppressed in the sum over all paths, whereas narrow ``flux
tubes" (Fig.9b) are favoured [Wil]. Note this argument was made within
the pure gauge theory, without explicitly including fermionic degrees of
freedom.

In the Hamiltonian formulation of the lattice theory due to Kogut and
Susskind, the (spatial) link holonomies play again an
important role [KogSus]. Together with an appropriate set of
conjugate momentum
variables they form a closed Poisson-bracket algebra, which is of the
form of a semi-direct product. This algebra is then quantized,
leading to a representation where the quantized link holonomies
$\hat U_l$ are diagonal. The wave functions in the Hilbert space are
not gauge-invariant, and the physical subspace has to be projected
out by imposing the Gauss law constraint as a Dirac condition on
quantum states. It turns out that gauge-invariant states can be
labelled by closed paths of links on the lattice.

However, with growing lattice size, selecting the physical subspace and
calculating the action of the Hamiltonian on it become quickly
involved. Furthermore this set of variables does
not seem particularly suited for treating the weak coupling limit. For
that reason, there have been various alternative proposals to formulate
the Hamiltonian lattice theory in an explicitly gauge-invariant
manner [FurKol, GaLeTr, RovSmo3, Br\"u]. They involve Hilbert spaces
of gauge-invariant states labelled by loops, and the action of the
Hamiltonian operator typically results in geometric deformations or
rearrangements (fusion, fission etc.) of these loop arguments. (A lattice
analogue of the equivalence theorems discussed in Sec.9 above is investigated
by Durhuus [Dur], more precisely, he derives a necessary and sufficient
condition on the gauge group $G$ such that the linear span of products of
Wilson loop variables is dense in the space of continuous gauge-invariant
observables.)

Unfortunately there
is now another drawback, since such bases of loop states are vastly
overcomplete (due to the Mandelstam constraints),
which renders the physical interpretation of the geometric
picture of ``loops interacting through the action of the Hamiltonian"
rather obscure. The problem is how to isolate in an efficient way
a set of independent loop states, since the number of loops on the
lattice grows very fast with growing lattice size. A frequently used
approximation employs a truncated loop basis, which only considers
states labelled by loops that are shorter than a given number $n$
of links [FurKol, Br\"u], and calculate the spectrum of the Hamiltonian
in this restricted context. This
seems to work reasonably well, at least for some sectors of the theory.
However, what one ideally would like to have is a formulation
directly in terms of the physical degrees of freedom, which is
gauge-invariant and has a minimal redundancy with respect to the
Mandelstam constraints. Such a set of variables has been given in
[Lol2], and recent results, obtained in a Lagrangian context, show that
it is indeed possible to perform calculations directly in terms of
these independent loop variables [Lol3].

Another advantage of the loop representation is that it can be naturally
extended to include fermions. Gauge-invariant variables may be obtained by
``gluing" fermion fields $\Psi$ to the ends of an open path $w$, leading
to the holonomy-dependent variables $\Psi^\dagger (y)U_w(y,x)\Psi(x)$. The
canonical loop algebra of the pure gauge theory (cf. the next section) can be
enlarged by these path variables, and quantized. The corresponding dynamics of
paths and loops on the lattice, together with some small scale calculations is
described by Gambini and Setaro [GamSet] (see also for references to the same
method applied to $U(1)$-gauge theory).

The lattice methods described above cannot be readily applied to the case of
general relativity, because the way in which the lattice approximation has
been set up explicitly violates diffeomorphism invariance. This is a serious
obstacle since the diffeomorphism group acts as a gauge group (not just as a
symmetry group like the Lorentz group, say), and a major problem in lattice
formulations of gravity has been to recover diffeomorphism invariance in
the continuum limit. Some problems that occur in translating the
$SL(2,\C)$-loop
approach \`a la Rovelli and Smolin to the lattice have been discussed by
Renteln [Ren]. - In spite of the difficulties of applying standard lattice
methods to the connection formulation of gravity,
the discrete structures appearing in its canonical quantization (cf.
Sec.15) indicate that lattice-like constructions may still play a role.
However, they would be defined through their intrinsic topology and
connectivity (i.e. diffeomorphism-invariant properties) rather than through
an imbedding as hypercubic lattices into a metric manifold.
For example, there are ways of defining diffeomorphism-invariant measures
on the space ${\cal A}/\cal G$ of connections modulo gauge transformations
introduced in Sec.5. They allow for the integration of connection
configurations that are supported on one-dimensional lattices or graphs
[AshLew2, Bae].

\vskip1.5cm

\line{\ch 14. Loop algebras\hfil}

In this section we will discuss some algebras of
loop- and path-dependent functions. The algebras we shall be concerned
with involve holonomies and are defined either on a configuration space
$\cal A$ or a phase space $T^*\cal A$. The first such algebra was
written down by Mandelstam in a Lagrangian framework [Man2]. Splitting
${\cal F}_{\mu\nu}(\g,x)={\cal F}_{\mu\nu}^a(\g,x)X_a$ (c.f.
(11.3)) into its components ${\cal F}_{ij}$ and ${\cal F}_{0i}$,
$i=1,2,3$, he finds for the equal-time commutators of an $SU(2)$ Yang-Mills
theory

$$
\eqalign{
&[{\cal F}_{ij}^a(\g,x), {\cal F}_{kl}^b(\g',x')]=0,\cr
&[{\cal F}_{0i}^a(\g,x), {\cal F}_{jk}^b(\g',x')]=
-i\d_{ab}(\d_{ik}\del_j -\d_{ij}\del_k)\d^3(x-x')
+i\e_{abc}\int_{\g'} d\xi_i\d^3(x-\xi){\cal F}_{jk}^c(\g',x'),\cr
&[{\cal F}_{0i}^a(\g,x), {\cal F}_{0j}^b(\g',x')]=
i\e_{abc}\int_{\g'}d\xi_i\d^3(x-\xi){\cal F}_{0j}^c(\g',x')
+i\e_{abc}\int_\g d\xi_j \d^3(x'-\xi){\cal F}_{0i}^c(\g,x).\cr}
\eqno(14.1)
$$

\ni The relations (14.1) define a closing algebra (if
we include the unit generator appearing on the right-hand side of
the second equation), i.e. the commutator of two ${\cal F}$'s is
again proportional to an algebra generator. A characteristic feature is the
fact that the structure constants of the  algebra are distributional
and depend on the path configurations appearing as arguments of the $\cal
F$-variables.

In the Hamiltonian formulation in terms of traced holonomies, one
finds a similar algebra structure after introducing conjugate holonomy
variables on the Yang-Mills phase space (in the $A_0=0$-gauge)
coordinatized by the pairs $(A_i^a(x),E_a^i(x))$, with canonical
Poisson brackets $\{A_i^a(x),E^j_b(y)\}=\d^a_b \d_i^j\d^3(x-y)$. They
depend on a loop $\g$, a marked point $\g(s)$ on $\g$, and both the
gauge potential and the generalized electric field and are defined as
([GamTri6, RovSmo2])

$$
T^i_{A,E}(\g,s):=\,{\rm Tr}\,U_\g(s,s)E^i(\g(s)).\eqno(14.2)
$$

\ni For the special case of SU(2), computing the Poisson brackets on
phase space of these loop variables leads to

$$
\eqalign{
&\{T(\g),T(\g')\}=0,\cr
&\{T^i(\g,s),T(\g')\}=
-\Delta^i(\g',\g(s))(T(\g\circ_s\g')-T(\g\circ_s\g'^{-1})),\cr
&\{T^i(\g,s),T^j(\g',t)\}=
-\Delta^i(\g',\g(s))(T^j(\g\circ_s\g',u(t))+
T^j(\g\circ_s\g'^{-1},u(t)))+\cr
&\hskip3.6cm +\Delta^j(\g,\g'(t))(T^i(\g'\circ_t\g,v(s))-
T^i(\g'\circ_t\g^{-1},v(s))).\cr}\eqno(14.3)
$$

\ni The structure constants $\Delta$ are again distributional:

$$
\Delta^i(\g,x)=\oint_{\g}dt\, \d^3(\g (t),x)\dot\g^i(t).
\eqno(14.4)
$$

\ni In its general form for gauge group SU(N), the algebra (14.3) was
first introduced by Gambini and Trias [GamTri6]. For the case of
$G=$SL(2,$\C$) (for which the algebra coincides with (14.3)), it was later
rediscovered by Rovelli and Smolin in a loop approach to canonical
quantum gravity [RovSmo2]. The same algebra is also relevant for
$2+1$-dimensional gravity, which may be formulated on a space $\cal A$ of
$su(1,1)$-valued connection one-forms [AHRSS, Smo1].
A generalization to a Hamiltonian algebra of fields depending on open
paths in the context of Higgs fields is described in [GaGiTr]. Related
loop-dependent algebras on the phase space of general relativity in terms
of the metric variables and that of scalar field theory are discussed by
Rayner [Ray1].

Note that the
algebra (14.3) has the form of a semi-direct product, with the abelian
subalgebra of the traced holonomies $\{T\}$ acted upon by the
non-abelian algebra of the $T^i$-variables, and is similar in
structure to the Lagrangian algebra (14.1).
This is not true for N$\not=2$, for which the
right-hand sides of the Poisson brackets in (14.3) are {\it not}
linear in $T$. In
order to make the algebra (14.3) non-distributional, one has to
``smear out" the loop variables  appropriately. One way of doing this
is to integrate (14.2) over a ribbon or ``strip" $R$ [Rov2, AshIsh1],
i.e. a non-degenerate one-parameter family $\g_t(s)=:R(s,t)$ of loops,
$t\in [0,1]$, according to

$$
T(R):=\int_0^1 dt\int_0^1 ds\,R'^i(s,t)\dot R^j(s,t) \e_{ijk}
T^k(\g_t(s)).\eqno(14.5)
$$

\ni The dot and the prime denote differentiation with respect to $s$ and
$t$ respectively, and $\e_{ijk}$ is the totally antisymmetric $\e$-tensor
in three dimensions. The resulting algebra of the loop and ribbon variables,
$T(\g)$ and $T(R)$, has real
structure constants that can be expressed in terms of the
intersection numbers of the loops and ribbons appearing in their
arguments. Although the algebraic structure of the relations (14.3)
can be neatly visualized by ``cutting and gluing" of diagrams of
loops and ribbons, a satisfactory physical interpretation
has not been given so far. Part of the problem is again the
overcompleteness of these non-local variables, which affects also the
ribbon variables $T(R)$. There have been attempts of integrating
the algebra (14.3) with the help of a formal group law expansion
[Lol1], and
thus possibly to obtain an new infinite-dimensional group structure on
the space of three-dimensional loops. Unfortunately it is difficult to
complete this program, again due to the lack of a suitable
topological and differentiable structure on loop space.

As has already been mentioned, the Wilson loop variables occurring in the
algebra (14.3) are not observables for general relativity, because they are
not invariant under diffeomorphisms. However, it is formally easy to implement
the three-dimensional spatial diffeomorphism invariance by requiring the
(generalized) Wilson loops $T$ to be constant on the orbits of the group
$Diff(\Sigma)$, as in (5.2). Since the canonical loop algebra
depends entirely on the topology of loop configurations, it projects down to
the quotient space $T^*{\cal A}/Diff({\Sigma})$.

Although it is not the subject of this article, let me point out that
the $2+1$-dimensional model for general relativity is an ideal testing ground
for some of the loop methods, since it has a non-abelian gauge group, but
still only a finite number of degrees of freedom. Loop algebras similar to
(14.3) have been used by Martin [Mar] and Nelson et al. [NelReg, NeReZe] to
describe and subsequently quantize $2+1$ gravity.

The main
utilization of the loop algebras described above is their importance
for canonical non-standard quantization schemes, which will be the
subject of the next section. (There are also straightforward lattice
analogues of the algebra (14.4), which have been employed in [RovSmo3,
Br\"u].) Another example of a non-trivial loop algebra, involving
non-intersecting loops, is due to `t Hooft [tHo1,2]. He supplements the
Wilson loops $T(\g)$ by a set of dual loop operators $\bar T(\g)$,
corresponding to magnetic flux lines. Both the $T$- and the $\bar
T$-operators commute among themselves, but the commutation relation
between a $T(\g)$ and a $T(\g')$ depends on the linking number of $\g$
and $\g'$. This algebra is used to extract qualitative behaviour of
different phases of Yang-Mills theory (see also the
discussions by Mandelstam [Man3] and Gambini and Trias [GamTri5] (who
construct a quantum representation of `t Hooft's algebra on a space of
loop states), and
the related work by Hosoya and Shigemoto [Hos, HosShi] on the idea of duality
between electric and magnetic flux lines).
\vskip1.5cm
\line{\ch 15. Canonical quantization\hfil}

The only non-perturbative quantization schemes put forward in the loop
formulation are Hamiltonian and operator-based. ``Non-perturbative" in this
context means ``not resorting to a perturbation expansion in terms of the
gauge potentials $A_\mu$".
One might argue that such attempts defeat the purpose of the
loop approach. Indeed, the hope in the non-local formulations we have
been discussing is often
for an {\it in}equivalence of the quantum theory and the usual
local field-theoretic quantization. Unfortunately, a corresponding, alternative
``loop perturbation theory" has not yet been developed. One also has to decide
about how to treat the Mandelstam constraints in the quantization, for example,
whether to solve them before or after the quantization, and different choices
may well lead to inequivalent quantum theories.

All existing canonical quantizations for Yang-Mills theory and gravity
in the loop
formulation postulate the existence of (self-adjoint) operator analogues
of a set of basic loop variables (such as the traced holonomies and
appropriate momentum variables), defined on some ``Hilbert space" of loop
functionals, such that their commutation relations are preserved in the
quantum theory. All of them are defined at a formal level, in the sense that
there is no proper Hilbert space structure, and the wave functions are just
elements of some linear function space, depending on loops. Some of them have
in
common that the action of the operator analogue $\hat T(\g)$ of the traced
holonomy in this function space is given by
multiplication by $T(\g)$. Note that we cannot
employ a Schr\"odinger-type quantization, because the algebra relations
of the basic
variables of the theory (for example, (14.3)) are not of the form of
canonical commutation relations.

Gambini and Trias were the first ones to write down an algebra of quantum
operators, realizing the algebra (14.3) [GamTri6]. Wave functions
in their approach
are labelled by individual loops and sets of loops, and there is a ``vacuum
state", the no-loop state $|0>$, which is annihilated by the momentum
operators $\hat T^i$. In this aspect their representation is similar to
the heighest-weight representation of the loop algebra proposed by Aldaya
and Navarro-Salas [AldNav].

For reasons inherent in the loop formulation of canonical gravity,
Rovelli and Smolin [RovSmo2] quantize an extended set of
SL(2,$\C$)-loop variables
$T^{i_1\dots i_n}$, which depend on $n$ ``electric field" variables
inserted into the traced holonomy, and are a straightforward generalization
of the expression (14.2). It turns out that in order to obtain a closed
Poisson algebra one has to include the infinite tower of these generalized
holonomy variables, for any $n$. The resulting algebra has a graded
structure, schematically given by

$$
\eqalign{
&\{ T^0,T^0\}=0,\cr
&\{ T^m,T^n\}\sim T^{m+n-1},\quad m+n>0\cr}\eqno(15.1)
$$

\ni where $m$ and $n$ denote the numbers of electric field insertions. This
algebra contains the algebra (14.3) as a closed subalgebra. The corresponding
quantum algebra obtained in [RovSmo2] is isomorphic to (15.1), with the
exception of higher-order correction terms proportional to $\hbar^k$,
$k\geq 2$, appearing on the right-hand sides of the commutators. (See also
the work by Rayner [Ray2] on this particular quantum representation.)

These higher-order terms do not appear in the quantization proposed in
[Lol1], where the semi-direct product structure of the algebra (14.3)
is exploited, using methods from the theory of unitary irreducible
representations
of semi-direct product algebras. This theory is very powerful in finite
dimensions, giving a complete classification and construction of such
representations. Some of the formalism can be applied to the
infinite-dimensional case too, although there is no reason to
expect that similarly strong results will hold.

Note at this stage that in spite of many similarities, ``solving" quantum
gauge theory and quantum gravity requires entirely different strategies.
In case of the former one needs some (generalized) measure on
a suitable space of loops (the domain space for
wave functions $\Psi$), and then looks for solutions $\Psi$ to the
eigenvector equation $\hat H_{YM}\Psi=E\Psi$, where $\hat H_{YM}$ is the
quantized Yang-Mills Hamiltonian. Up to now one does not know how to
tackle this problem non-perturbatively in the continuum theory, but progress
has been made in the lattice-regularized theory, as explained in Sec.13.

For the case of gravity, the quantum dynamics is contained in the constraint
equation $\hat H_{GR}\Psi=0$, i.e. physical wave functions $\Psi$ are
annihilated by the gravitational quantum Hamiltonian (they also have to be
annihilated by the quantum constraints corresponding to spatial
diffeomorphisms). Then, a suitable measure has to be found on the space
$\{\Psi\}$ of these solutions. It is not my intention to describe in any
detail the progress that has been made in finding such ``non-perturbative
solutions to quantum gravity" with the help of loop methods, because a
number of excellent reviews on the subject is already available [Rov1,2, Pul,
Smo2, Ash2].

There are many intriguing aspects to this line of research, and the
development of appropriate mathematical structures is not keeping pace with
the many heuristic ideas that are being put forward. Examples are the
relation of the quantum solutions to knot and link invariants [BrGaPu,
DiGaGr, AshLew2], the construction of diffeomorphism-invariant observables
[Smo3], and the introduction of a ``weave" of loops to model space at small
scales [AsRoSm, Ash2]. It is clear however that discrete mathematical
objects (like the generalized knot classes [RovSmo1]) play a prominent role.

One interesting mathematical tool that has been introduced to relate the
quantum representations on spaces of connection wave functions $\Psi([A])$
and loop wave functions $\tilde\Psi (\g)$ is the so-called loop transform.
The idea of exploiting the (non-linear) duality between those two spaces
appeared first in [GamTri5].
In the form introduced by Rovelli and Smolin [RovSmo2], the loop
transform reads

$$
\tilde\Psi(\g)=\int_{{\cal A}/{\cal G}}[dA]_{\cal G}\;T_A(\g)\Psi([A]),
\eqno(15.2)
$$

\ni where the Wilson loops $T_A(\g)$ serve as kernel of the transform. Also
self-adjoint operators may be translated from the connection to the loop
representation with the help of (15.2). For Yang-Mills theory, this
construction is of limited practical use, in lack of a well-defined
gauge-invariant measure $[dA]_{\cal G}$. Nevertheless many of its properties
have been explored in both gauge and gravitational theories (see, for
example, [Br\"uPul, AshIsh1, AshRov, Lol6, AshLew2, AshLol].

The approach that leads closest to the construction of a rigorous
representation theory of the loop algebra is that of Ashtekar and Isham
[AshIsh1], subsequently extended by Ashtekar and Lewandowski [AshLew2].
They start from the abelian subalgebra of the traced holonomies
$T(\g)$, i.e. the first line of the algebra (14.3), endow it with the
structure of a C$^*$-algebra, and then look for its
cyclic representations, using Gel'fand spectral theory. The Hilbert spaces
involved are given by spaces of square-integrable functions on the
space of ideals of the C$^*$-algebra, a certain completion $\overline{{\cal A}/
{\cal G}}$ of the space ${\cal A}/\cal G$, allowing also for distributional
connections $A$. The precise mathematical structure of the space
$\overline{{\cal A}/{\cal G}}$ and the inclusion of appropriate conjugate
momentum variables are currently under investigation.

Beyond these mainly kinematical considerations of how to quantize algebras
of basic phase space variables, little is known about a proper
formulation of the Hamiltonian dynamics, i.e. about how to express
the quantum Hamiltonians of Yang-Mills theory and gravity as well-defined
(self-adjoint) operators in terms
of the non-local
(quantized) loop variables. One main ingredient is the choice of an
appropriate regularization and renormalization prescription (which in the
case of gravity has to respect diffeomorphism invariance).
On the other hand, we know from
finite-dimensional examples
that the quantization of non-canonical commutation relations at a kinematic
level usually is
non-unique, and we expect the situation to be much worse in the present,
field-theoretic case (see also [AshIsh2] on the ambiguity of field-theoretic
quantizations). Whichever representation theory we come up with for the
loop algebra, further physical criteria will be needed to decide which of
the multitude of possible representations is physically relevant, for example,
by selecting those in which the Hamiltonian assumes a simple form, and by
finding ways to relate them to physical observations.

\vfill\eject

\line{\ch References\hfil}
\vskip0.8cm

\item{[Ada]} Adams, J.F.: Infinite loop spaces, Princeton University
  Press, 1978

\item{[AldNav]} Aldaya, V. and Navarro-Salas, J.: New solutions of the
  Hamiltonian and diffeomorphism constraints of quantum gravity from a
  highest weight loop representation, Phys. Lett. 259B (1991) 249-255

\item{[AmbSin]} Ambrose, W. and Singer, I.M.: A theorem on holonomy,
  Trans. Amer. Math. Soc. 75 (1953) 428-443

\item{[Ana1]} Anandan, J.: Holonomy groups in gravity and gauge fields,
  in Proceedings of the Conference on Differential Geometric Methods in
  Theoretical Physics, Trieste 1981, eds. G. Denardo and H.D. Doebner,
  World Scientific, 1983

\item{[Ana2]} Anandan, J.: Gauge fields, quantum interference, and holonomy
  transformations, Phys. Rev. D33 (1986) 2280-2287

\item{[Ana3]} Anandan, J.: Remarks concerning the geometries of gravity and
  gauge fields, to appear in Directions in General Relativity, Vol.1, ed. B.L.
  Hu et al., Cambridge University Press, 1993

\item{[Are1]} Aref'eva, I.Ya.: The gauge field as chiral field on the path
  and its integrability, Lett. Math. Phys. 3 (1979) 241-247

\item{[Are2]} Aref'eva, I.Ya.: Non-abelian Stokes formula, Teor. Mat. Fiz. 43
  (1980) 111-116

\item{[Are3]} Aref'eva, I.Ya.: Quantum contour field equations, Phys. Lett.
  93B (1980) 347-353

\item{[Are4]} Aref'eva, I.Ya.: The integral formulation of gauge
  theories - strings, bags or something else, Lectures given at the
  17th Karpacz Winter School, 1980

\item{[Are5]} Aref'eva, I.Ya.: Elimination of divergences in the integral
  formulation of the Yang-Mills theory, Pis'ma Zh. Eksp. Teor. Fiz. 31
  (1980) 421-425

\item{[Ash1]} Ashtekar, A.: Physics in loop space, Cochin lecture notes
  prepared by R.S. Tate, in: Quantum gravity, gravitational radiation and
  large scale structure in the universe, ed. B.R. Iyer, S.V. Dhurandhar and
  K. Babu Joseph, 1993

\item{[Ash2]} Ashtekar, A.: Mathematical problems of non-perturbative
  quantum general relativity, in: Proceedings of the 1992 Les Houches
  summer school on gravitation and quantization, ed. B. Julia,
  North-Holland, Amsterdam, 1993

\item{[AHRSS]} Ashtekar, A., Husain, V., Rovelli, C., Samuel, J. and Smolin,
  L.: 2+1 gravity as a toy model for the 3+1 theory, Class. Quan. Grav. 6
  (1989) L185-193

\item{[AshIsh1]} Ashtekar, A. and Isham, C.J.: Representations of the
  holonomy algebras of gravity and non-Abelian gauge theories, Class.
  Quan. Grav. 9 (1992) 1433-1467

\item{[AshIsh2]} Ashtekar, A. and Isham, C.J.: Inequivalent observable
  algebras: another ambiguity in field quantisation, Phys. Lett. 274B
  (1992) 393-398

\item{[AshLew1]} Ashtekar, A. and Lewandowski, J.: Completeness of Wilson
  loop functionals on the moduli space of $SL(2,\C)$ and
  $SU(1,1)$-connections, Class. Quan. Grav. 10 (1993) L69-74

\item{[AshLew2]} Ashtekar, A. and Lewandowski, J.: Representation theory of
  analytic holonomy $C^*$-algebras,
  to appear in: Knot theory and quantum gravity, ed. J. Baez,
  Oxford University Press

\item{[AshLol]} Ashtekar, A. and Loll, R.: A new loop transform for 2+1
  gravity, in preparation

\item{[AshRov]} Ashtekar, A. and Rovelli, C.: A loop representation for the
  quantum Maxwell field, Class. Quan. Grav. 9 (1992) 1121-1150

\item{[AsRoSm]} Ashtekar, A., Rovelli, C. and Smolin, L.: Weaving a classical
  metric with quantum threads, Phys. Rev. Lett. 69 (1992) 237-240

\item{[BGNS]} Brandt, R.A., Gocksch, A., Neri, F. and Sato, M.-A.: Loop
  space, Phys. Rev. D12 (1982) 3611-3640

\item{[Bae]} Baez, J.C.: Diffeomorphism-invariant generalized measures on the
  space of connections modulo gauge transformations, to appear in Proceedings
  of the Conference on Quantum Topology, Manhattan, Kansas, 1993

\item{[Bar]} Barrett, J.W.: Holonomy and path structures in general
  relativity and Yang-Mills theory, Int. J. Theor. Phys. 30 (1991)
  1171-1215

\item{[BerUrr]} Berenstein, D.E. and Urrutia, L.F.: The relation between
  the Mandelstam and the Cayley-Hamilton identities, preprint Bogot\`a and
  M\'exico, June 1993

\item{[Bia]} Bialynicki-Birula, I.: Gauge-invariant variables in the
  Yang-Mills theory, Bull. Acad. Polon. Sci. 11 (1963) 135-138

\item{[Ble]} Blencowe, M.P.: The Hamiltonian constraint in quantum gravity,
  Nucl. Phys. B341 (1990) 213-251

\item{[BrGaPu]} Br\"ugmann, B., Gambini, R. and Pullin, J.: Knot
  invariants as nondegenerate quantum geometries, Phys. Rev. Lett.
  68 (1992) 431-434

\item{[BrNeSa]} Brandt, R.A., Neri, F. and Sato, M.-A.: Renormalization of
  loop functions for all loops, Phys. Rev. D24 (1981) 879-902

\item{[Bra]} Brali\'c, N.E.: Exact computation of loop averages in
  two-dimensional Yang-Mills theory, Phys. Rev. D22 (1980) 3090-3103

\item{[Br\"u]} Br\"ugmann, B.: The method of loops applied to lattice
  gauge theory, Phys. Rev. D43 (1991) 566-579

\item{[Br\"uPul]} Br\"ugmann, B. and Pullin, J.: On the constraints of
  quantum gravity in the loop representation, Nucl. Phys. B390 (1993)
  399-438

\item{[Bry]} Brylinski, J.M.: The Kaehler geometry of the space of knots
  in a smooth threefold, Penn State pure mathematics report No. PM93
  (1990)

 \item{[ChoDeW]} Choquet-Bruhat, Y. and DeWitt-Morette, C. (with
  M. Dillard-Bleick): Analysis, manifolds and physics, revised edition,
  North-Holland, Amsterdam, 1982

\item{[CoqPil]} Coquereaux, R. and Pilch, K.: String structures on loop
  bundles, Comm. Math. Phys. 120 (1989) 353-378

\item{[CorHas]} Corrigan, E. and Hasslacher, B.: A functional equation for
  exponential loop integrals in gauge theories, Phys. Lett. 81B (1979)
  181-184

\item{[DiGaGr]} Di Bartolo, C., Gambini, R. and Griego, J.: The extended
  loop group: an infinite dimensional manifold associated with the loop space,
  preprint Montevideo, June 1992

\item{[DotVer]} Dotsenko, V.S. and Vergeles, S.N.: Renormalizability of
  phase factors in non-abelian gauge theory, Nucl. Phys. B169 (1980)
  527-546

\item{[Dur]} Durhuus, B.: On the structure of gauge-invariant classical
  observables in lattice gauge theories, Lett. Math. Phys. 4 (1980)
  515-522

\item{[DurLei]} Durhuus, B. and Leinaas, J.M.: On the loop space formulation
  of gauge theories, Physica Scripta 25 (1982) 504-510

\item{[DurOle]} Durhuus, B. and Olesen, P.: Eigenvalues of the Wilson
  operator in multicolor QCD, Nucl. Phys. B184 (1981) 406-428

\item{[Fis]} Fischer, A.E.: A grand superspace for unified field
  theories, Gen. Rel. Gravit. 18 (1986) 597-608

\item{[FiGaKa]} Fishbane, P.M., Gasiorowicz, S. and Kaus, P.: Stokes's
  theorem for non-abelian fields, Phys. Rev. D 24 (1981) 2324-2329

\item{[ForGam]} Fort, H. and Gambini, R.: Lattice QED with
  light fermions in
  P representation, Phys. Rev. D44 (1991) 1257-1262

\item{[Fr\"o]} Fr\"ohlich, J.: Some results and comments on quantized gauge
  fields, in: Recent developments in gauge theories, Carg\`ese 1979
  (eds. `t Hooft et al)

\item{[Ful1]} Fulp, R.O.: The nonintegrable phase factor and gauge theory,
  to be published in Proceedings of the 1990 Summer Institute on Differential
  Geometry, Symposia in Pure Mathematics Series

\item{[Ful2]} Fulp, R.O.: Connections on the path bundle of a principal
  fibre bundle, Math. preprint North Carolina State University

\item{[FurKol]} Furmanski, W. and Kolawa, A.: Yang-Mills vacuum: An attempt
  at lattice loop calculus, Nucl. Phys. B291 (1987) 594-628

\item{[GaGiTr]} Gambini, R., Gianvittorio, R. and Trias, A.: Gauge Higgs
  dynamics in the loop space, Phys. Rev. D38 (1988) 702-705

\item{[GaLeTr]} Gambini, R., Leal, L. and Trias, A.: Loop
  calculus for lattice gauge theories, Phys. Rev. D39 (1989) 3127-3135

\item{[Gam]} Gambini, R.: Loop space representation of quantum general
  relativity and the group of loops, Phys. Lett. 255B (1991) 180-188

\item{[GamGri]} Gambini, R. and Griego, J.: A geometric approach to the
  Makeenko-Migdal equations, Phys. Lett. 256B (1991) 437-441

\item{[GamSet]} Gambini, R. and Setaro, L.: SU(2) QCD in the path
  representation, preprint Montevideo, April 1993

\item{[GamTri1]} Gambini. R. and Trias, A.: Path-dependent formulation of
  gauge theories and the origin of field copies in the non-Abelian case,
  Phys. Rev. D21 (1980) 3413-3416

\item{[GamTri2]} Gambini, R. and Trias, A.: Second quantization
  of the free electromagnetic field as quantum mechanics in the loop
  space, Phys. Rev. D22 (1980) 1380-1384

\item{[GamTri3]} Gambini, R. and Trias, A.: Geometrical origin of
  gauge theories, Phys. Rev. D23 (1981) 553-555

\item{[GamTri4]} Gambini, R. and Trias, A.: Chiral formulation of Yang-Mills
  equations: A geometric approach, Phys. Rev. D27 (1983) 2935-2939

\item{[GamTri5]} Gambini, R. and Trias, A.: On confinement in pure
  Yang-Mills theory, Phys. Lett. 141B (1984) 403-406

\item{[GamTri6]} Gambini, R. and Trias, A.: Gauge dynamics in the
  C-representation, Nucl. Phys. B278 (1986) 436-448

\item{[GerNev1]} Gervais, J.-L. and Neveu, A.: The quantum dual string
  functional in Yang-Mills theory, Phys. Lett. 80B (1979) 255-258

\item{[GerNev2]} Gervais, J.-L. and Neveu, A.: Local harmonicity of the
  Wilson loop integral in classical Yang-Mills theory, Nucl. Phys. B153
  (1979) 445-454

\item{[Gil]} Giles, R.: Reconstruction of gauge potentials from Wilson loops,
  Phys. Rev. D24 (1981) 2160-2168

\item{[GliVir]} Gliozzi, F. and Virasoro, M.A.: The interaction among dual
  strings as a manifestation of the gauge group, Nucl. Phys. B164 (1980)
  141-151

\item{[GoLeSt]} Goldberg, J.N., Lewandowski, J. and Stornaiolo, C.:
  Degeneracy in loop variables, Comm. Math. Phys. 148 (1992) 377-402

\item{[Gro]} Gross, L.: A Poincar\'e Lemma for connection forms,
  J. Funct. Anal. 63 (1985) 1-46

\item{[Gu]} Gu, C.-H.: On classical Yang-Mills fields, Phys. Rep. 80
  (1981) 251-337

\item{[GuWan]} Gu, C.-H. and Wang, L.-L.Ch.: Loop-space formulation of gauge
  theories, Phys. Rev. Lett. 25 (1980) 2004-2007

\item{[HoScTs]} Hong-Mo, C., Scharbach, P. and Tsun, T.S.: On loop
  space formulation of gauge theories, Ann. Phys. (NY) 166 (1986)
  396-421

\item{[HonTsu]} Hong-Mo, C. and Tsun, T.S.: Gauge theories in loop
  space, Acta Phys. Pol. B17 (1986) 259-276

\item{[Hos]} Hosoya, A.: Duality for the Lorentz force in loop space,
  Phys. Lett. 92B (1980) 331-332, Err.:96B (1980) 444

\item{[HosShi]} Hosoya, A. and Shigemoto, K.: Dual potential and magnetic
  loop operator, Prog. Theor. Phys. 65 (1981) 2008-2022

\item{[Ish]} Isham, C.J.: Loop algebras and canonical quantum gravity, in:
  Contemporary Mathematics, vol.132, ed. M. Gotay, V. Moncrief and J. Marsden,
  American Mathematical Society, Providence, 1992

\item{[Kau1]} Kauffman, L.H.: On knots, Princeton University Press, 1987

\item{[Kau2]} Kauffman, L.H.: Knots and Physics, World Scientific,
  Singapore, 1991

\item{[KobNom]} Kobayashi, S. and Nomizu, K.: Foundations of differential
  geometry, Vol.1, Interscience, New York, 1969

\item{[KogSus]} Kogut, J. and Susskind, L.: Hamiltonian formulation of
  Wilson's lattice gauge theories, Phys. Rev. D11 (1975) 395-408

\item{[Kuc]} Kucha\v r, K.V.: Canonical quantum gravity, in: Proceedings of the
  13th International Conference on General Relativity and Gravitation, ed.
  C. Kozameh, IOP Publishing, Bristol, 1993

\item{[Lew]} Lewandowski, J.: Group of loops, holonomy maps, path bundle and
  path connection, Class. Quan. Grav. 10 (1993) 879-904

\item{[Lic]} Lichnerowicz, A.: Global theory of connections and holonomy
  groups, Noordhoff International Publishing, 1976 (French edition
  published in 1955)

\item{[Lip]} Lipschutz, S.: General topology, McGraw-Hill, New York, 1965

\item{[Lol1]} Loll, R.: A new quantum representation for canonical gravity and
  SU(2) Yang-Mills theory, Nucl. Phys. B350 (1991) 831-860

\item{[Lol2]} Loll, R.: Independent SU(2)-loop variables and the reduced
  configuration space of SU(2)-lattice gauge theory, Nucl. Phys.
  B368 (1992) 121-142

\item{[Lol3]} Loll, R.: Yang-Mills theory without Mandelstam constraints,
  Nucl. Phys. B400 (1993) 126-144

\item{[Lol4]} Loll, R.: Lattice gauge theory in terms of independent Wilson
  loops, Nucl. Phys. B, Proc. Suppl. 30 (1993) 224-227

\item{[Lol5]} Loll, R.: Loop variable inequalities in gravity and gauge
  theory, Class. Quan. Grav., to appear

\item{[Lol6]} Loll, R.: Loop formulation of gauge theory and gravity,
  to appear in: Knots and Quantum Gravity, ed. J. Baez, Oxford University
  Press

\item{[Lol7]} Loll, R.: Loop approaches to gauge field theories, Teor. Mat.
  Fiz. 93 (1992) 481-505

\item{[MakMig1]} Makeenko, Yu.M. and Migdal, A.A.: Exact equation for the
  loop average in multicolor QCD, Phys. Lett. 88B (1979) 135-137 (E: 89B
  (1980) 437)

\item{[MakMig2]} Makeenko, Yu.M. and Migdal, A.A.: Quantum chromodynamics as
  dynamics of loops, Nucl. Phys. B188 (1981) 269-316

\item{[Man1]} Mandelstam, S.: Quantum electrodynamics without potentials,
  Ann. Phys. (NY) 19 (1962) 1-24

\item{[Man2]} Mandelstam, S.: Feynman rules for electromagnetic and Yang-Mills
  fields from the gauge-independent field-theoretic formalism, Phys. Rev.
  175 (1968) 1580-1603

\item{[Man3]} Mandelstam, S.: Charge-monopole duality and the phases of
  non-Abelian gauge theories, Phys. Rev. D19 (1979) 2391-2409

\item{[Mar]} Martin, S.P.: Observables in 2+1 dimensional gravity, Nucl. Phys.
  B327 (1989) 178-204

\item{[Men1]} Mensky, M.B.: Group of parallel transports and description
  of particles in curved space-time, Lett. Math. Phys. 2 (1978) 175-180

\item{[Men2]} Mensky, M.B.: Application of the group of paths to the
  gauge theory and quarks, Lett. Math. Phys. 3 (1979) 513-520

\item{[Men3]} Mensky, M.B.: The group of paths (in Russian), Nauka, Moscow,
  1983

\item{[Mic]} Michor, P.W.: Manifolds of differentiable mappings, Shiva
  Publishing Limited, Orpington, 1980

\item{[Mig1]} Migdal, A.A.: Properties of the loop average in QCD,
  Ann. Phys. (NY) 126 (1980) 279-290

\item{[Mig2]} Migdal, A.A.: Loop equations and 1/N expansion, Phys. Rep. 102
  (1983) 199-290

\item{[MosShn]} Mostow, M.A. and Shnider, S.: Does a generic connection
  depend continuously on the curvature?, Commun. Math. Phys. 90 (1983) 417-432

\item{[Nam]} Nambu, Y.: QCD and the string model, Phys. Lett. 80B (1979)
  372-376

\item{[NelReg]} Nelson, R. and Regge, T.: Homotopy groups and 2+1 dimensional
  quantum gravity, Nucl. Phys. B328 (1989) 190-202

\item{[NeReZe]} Nelson, J., Regge, T. and Zertuche, F.: Homotopy groups and
  (2+1)-dimensional quantum De Sitter gravity, Nucl. Phys. B339 (1990)
  516-532

\item{[Pol1]} Polyakov, A.M.: String representations and hidden symmetries
  for gauge fields, Phys. Lett. 82B (1979) 247-250

\item{[Pol2]} Polyakov, A.M.: Gauge fields as rings of glue, Nucl. Phys. B164
  (1979) 171-188

\item{[Pol3]} Polyakov, A.M.: Gauge fields and strings, Harwood Academic
  Publishers, 1987

\item{[Pul]} Pullin, J.: Knot theory and quantum gravity in loop space: a
  primer, in: Proceedings of the Vth Mexican School of Particles and Fields,
  ed. J.L.Lucio, World Scientific, Singapore, 1993

\item{[PreSeg]} Pressley, A. and Segal, G.: Loop groups, Clarendon Press,
  Oxford, 1986

\item {[Ray1]} Rayner, D.: A formalism for quantising general relativity
  using non-local variables, Class. Quan. Grav. 7 (1990) 111-134

\item{[Ray2]} Rayner, D.: Hermitian operators on quantum general
  relativity loop space, Class. Quan. Grav. 7 (1990) 651-661

\item{[Ren]} Renteln, P.: Some results of SU(2) spinorial lattice gravity,
  Class. Quan. Grav. 7 (1990) 493-502

\item{[Rov1]} Rovelli, C.: Holonomies and loop representation in quantum
  gravity, in: The Newman Festschrift, ed. A. Janis and J. Porter,
  Birkh\"auser, Boston, 1991

\item{[Rov2]} Rovelli, C.: Ashtekar formulation of general relativity and
  loop-space non-perturbative quantum gravity: A report, Class. Quan.
  Grav. (1991) 1613-1675

\item{[RovSmo1]} Rovelli, C. and Smolin, L.: Knot theory and quantum
  gravity, Phys. Rev. Lett. 61 (1988) 1155-1158

\item{[RovSmo2]} Rovelli, C. and Smolin, L.: Loop space representation of
  quantum general relativity, Nucl. Phys. B331 (1990) 80-152

\item{[RovSmo3]} Rovelli, C. and Smolin, L.: Loop representation for lattice
  gauge theory, preprint Pittsburgh and Syracuse 1990

\item{[Sch1]} Sch\"aper, U.: Geometry of loop spaces, I. A Kaluza-Klein type
  point of view, preprint Freiburg THEP 91/3, March 1991, 41pp.

\item{[Sch2]} Sch\"aper, U.: Geodesics on loop spaces, preprint Freiburg
  THEP 92/12, March 1992, 6pp.

\item{[Smo1]} Smolin, L.: Loop representation for quantum gravity in 2+1
  dimensions, in: Proceedings of the 12th Johns Hopkins Workshop on
  Knots, Topology and Quantum Field Theory, ed. L. Lusanna, World Scientific,
  Singapore, 1990

\item{[Smo2]} Smolin, L.: Recent developments in nonperturbative quantum
  gravity, in: Proceedings of the XXII GIFT International Seminar on
  Theoretical Physics, Quantum Gravity and Cosmology, World Scientific,
  Singapore, 1992

\item{[Smo3]} Smolin, L.: Finite diffeomorphism invariant observables in
  quantum gravity, preprint Syracuse SU-GP-93/1-1

\item{[Sta]} Stasheff, J.: Differential graded Lie algebras, Quasi-Hopf
  algebras and higher homotopy algebras, preprint UNC-MATH-91-3

\item{[Tav]} Tavares, J.N.: Chen integrals, generalized loops and
  loop calculus, math preprint, University of Porto (1993)

\item{[Tel]} Teleman, M.C.: Sur les connexions infinit\'esimales qu'on peut
  d\'efinir dans les structures fibr\'ees diff\'erentiables de base donn\'ee,
  Ann. di Mat. Pura ed Appl. 62 (1963) 379-412

\item{[tHo1]} 't Hooft, G.: On the phase transition towards permanent quark
  confinement, Nucl. Phys. B138 (1978) 1-25

\item{[tHo2]} 't Hooft, G.: A property of electric and magnetic flux in
  non-Abelian gauge theories, Nucl. Phys. B153 (1979) 141-160

\item{[Wil]} Wilson, K.: Confinement of quarks, Phys. Rev. D10 (1974)
  2445-2459

\item{[WuYan]} Wu, T.T. and Yang, C.N.: Concept of non-integrable phase
  factors and global formulation of gauge theories, Phys. Rev. D12 (1975)
  3845-3857

\item{[Yan]} Yang, C.N.: Integral formalism for gauge fields, Phys. Rev.
  Lett. 33 (1974) 445-447

\end